\documentclass[uplatex,dvipdfmx]{pasj01}
\draft
\usepackage{bm}
\usepackage[hang,small,bf]{caption}
\usepackage[subrefformat=parens]{subcaption}
\usepackage{graphicx}
\usepackage{color}
\usepackage{ulem}
\begin{document}


\title{
Thermal emission from the amorphous dust: 
An alternative possibility of the origin of the anomalous microwave emission
}

\author{
Masashi \textsc{nashimoto},\altaffilmark{1,2,}$^{*}$
Makoto \textsc{hattori},\altaffilmark{1}
Ricardo \textsc{g{\'e}nova-santos},\altaffilmark{3,4}
Fr{\'e}d{\'e}rick \textsc{poidevin}\altaffilmark{3,4}
}
\email{m.nashimoto@astr.tohoku.ac.jp}

\altaffiltext{1}
{Astronomical Institute, Tohoku University, 6-3, Aramaki-Aza-Aoba, Aoba-ku, Sendai, Miyagi, Japan}
\altaffiltext{2}
{Graduate Program on Physics for the Universe (GP-PU), Tohoku University, 6-3, Aramaki-Aza-Aoba, Aoba-ku, Sendai, Miyagi, Japan}
\altaffiltext{3}
{Instituto de Astrofis{\'i}ca de Canarias, E-38200 La Laguna, Tenerife, Canary Islands, Spain}
\altaffiltext{4}
{Departamento de Astrof{\'i}sica, Universidad de La Laguna (ULL), E-38206 La Laguna, Tenerife, Spain}


\maketitle

\begin{abstract}
Complete studies of the radiative processes of thermal emission from the amorphous dust from microwave through far infrared wavebands are presented by taking into account, self-consistently for the first time, the standard two-level systems (TLS) model of amorphous materials.
The observed spectral energy distributions (SEDs) for the Perseus molecular cloud (MC) and W43 from microwave through far infrared are fitted with the SEDs calculated with the TLS model of amorphous silicate. 
We have found that the model SEDs well reproduce the observed properties of the anomalous microwave emission (AME). 
The present result suggests an alternative interpretation for the AME being carried by the resonance emission of the TLS of amorphous materials without introducing new species.
Simultaneous fitting of the intensity and polarization SEDs for the Perseus MC and W43 are also performed.
The amorphous model reproduces the overall observed feature of the intensity and polarization SEDs of the Perseus MC and W43.
However, the model's predicted polarization fraction of the AME is slightly higher than the QUIJOTE upper limits in several frequency bands.
A possible improvement of our model to resolve this problem is proposed.
Our model predicts that interstellar dust is amorphous materials having very different physical characteristics compared with terrestrial amorphous materials.
\end{abstract}

\section{Introduction}
\label{sec:intro}

Studies of the physical processes of thermal emission from Galactic dust have been a long-standing problem and are still of significance.
The typical temperature of Galactic interstellar dust is about 20 K (Schlegel et al. \yearcite{Schlegel+1998}; \cite{PlanckXI_2014}).
Its emission appears predominantly at long wavelengths from the far infrared through microwave.
Since the whole sky is covered by the emission from Galactic interstellar dust, thermal emission from the Galactic dust is a serious obstacle for the detection of B-mode polarization signals from cosmic microwave background (CMB) radiation imprinted by primordial gravitational waves.
The success of CMB B-mode polarization observations relies on how accurately we can remove the Galactic dust signals from observational data.
To tackle this difficult task, many CMB experiments are under way and others are being planned
(e.g.
ACTPol \citep{ACTPol_2014};
BICEP2/3 and the Keck Array \citep{BICEP3_2016};
CLASS \citep{CLASS_2014};
GroundBIRD \citep{GroundBIRD_2016};
LiteBIRD \citep{LiteBIRD_2014};
PIPER \citep{PIPER_2016};
POLARBEAR and the Simons Array \citep{SA_2014};
QUIJOTE \citep{QUIJOTE_2012};
the Simons Observatory \citep{SO_2019};
SPIDER \citep{SPIDER_2018};
SPTPol \citep{SPTPol_2012}
).
These surveys provide extremely high precision data on the microwave sky with wide sky coverage in many different wavebands.
It is certain that significant progress in our understanding of interstellar dust will be made with these data.
Therefore, theoretical studies of the physical processes of thermal emission from Galactic dust must be undertaken now to achieve fruitful outcomes from these data.

The origin of anomalous microwave emission, which is abbreviated AME found ubiquitously in the Galaxy at around 10--30 GHz (e.g. see \citet{Dickinson_2013} for a summary of AME observations in HII regions) is still under debate.
The spatial correlation between AME and Galactic interstellar dust strongly indicates that AME originates from a kind of dust \citep{Davis+2006}.
The most popular model of the origin of AME is electric dipole emission radiated by charged rotating dust with a frequency of several tens of GHz, as proposed by \citet{Draine+1998}; this is referred to as the spinning dust model.
A carrier of the spinning dust is supposed to be very small grains producing rotation at ultra high frequencies. 
The fact of the lack of AME in cold dense cores supports the spinning dust origin of AME since the lack of the small grains is expected in dense clouds \citep{Tibbs+2016}.
Polycyclic aromatic hydrocarbon (PAH) has been proposed as one of the plausible candidates for spinning dust \citep{Draine+1998}.
However, no observational correlation between amount of PAH and the intensity of AME, as reported by \citet{Hensley+2016}, contradicts the PAH possibility.
A new species of very small dust grains named nanosilicates has been introduced as another possible carrier of the spinning dust (Hoang et al. \yearcite{Hoang+2016}; \cite{Hensley+2017}).
The problem with this possibility is that up to now, apart from AME, no signature to confirm the existence of the nanosilicate has been observed.
The nanosilicate is only observable as AME.
Therefore, it is hard to check whether the carrier of the spinning dust is such a new family of dust grains or not. 
Magnetic dust emission has been proposed as another candidate for the AME mechanism \citep{Draine+1999}.
The spins of electrons inside a magnetic dust grain align spontaneously to settle down to the minimum energy state.
Alignment is disturbed by thermal fluctuation.
Owing to the magnetic relaxation, the disturbed state tries to return to the original minimum energy state.
In course of this transition, microwave radiation is emitted.
This emission could be the origin of AME if the interstellar dust is magnetic \citep{Draine+2013}.
The magnetic dust emission model predicts a positive correlation between the temperature and intensity of AME.
However, \citet{Hensley+2016} found a negative correlation between the AME temperature and intensity that contradicts the predictions of the magnetic dust emission model.
A comprehensive review on the state of research of AME  is given by \citet{Dickinson+2018}.

Crucial clue to distinguishing the emission mechanisms of AME is offered by polarization observations.
\citet{Draine+2016} showed that the quantum effect suppresses the thermalization of the grain rotational kinetic energy of the spinning dust.
As a result, the alignment of grains is suppressed and the spinning dust model predicts a very low degree of polarization.
In the magnetic dust emission model, the high degree of AME polarization  is expected since the main carrier of magnetic dust emission is large grains which are aligned by the interstellar magnetic field.
Although the progress of AME polarization observations have been made by several projects (e.g. WMAP and QUIJOTE), there has as yet been no definite report of the detection of AME polarization (\cite{QUIJOTE1_2015}, \yearcite{QUIJOTE2_2017}).
The detection of polarization from AME has been reported for W43, but whether the reported polarization is a residual of the synchrotron emission of Galactic interstellar matters around W43 is still being debated.
Current observational upper limits somehow rule out the magnetic dust hypothesis, which typically predicts a higher polarization fraction.

Almost all types of interstellar dust are supposed to be made of amorphous materials.
For example, the broad emission line observed ubiquitously in interstellar space at 9.7 $\micron$ is considered to be a signature that one of the main components of interstellar dust is an amorphous silicate (\cite{Kraetschmer+1979}; \cite{Li+2001}).
Moreover, laboratory simulations of cosmic dust analogues suggest that various forms of amorphous carbon grains are more favorable than graphite grains (\cite{Colangeli+1995}; \cite{Zubko+1996}).
The observed spectrum of interstellar dust emission in submillimeter wavebands obtained by the Planck satellite is flatter than the spectrum expected from crystal dust (\cite{PlanckXI_2014}).
This is further evidence indicating that interstellar dust is composed of amorphous dust.
According to recent laboratory measurements of the emissivity of amorphous material, the emissivity of the amorphous material has complex frequency dependences that cannot be approximated by a single power law at longer than far infrared wavelengths (e.g., \cite{Coupeaud_2011}).
Physical diagnostics of the amorphous material appear in heat capacity and heat conductivity at very low temperature.
\citet{Zeller+1971} found that the heat capacity of the amorphous material below 1 K shows significant deviation from the Debye model and depends linearly on temperature instead of the cube of the temperature.
They also found that heat conductivity below 1 K is in excess of that expected for crystals and depends on the square of the temperature.
It has also been shown that these characteristics appear universally in any amorphous material.
This universality indicates that above-mentioned diagnostics observed in amorphous materials are governed by universal physics.
\citet{Anderson+1972} and \citet{Phillips_1972} independently proposed that heat absorption and heat transport by two-level systems from amorphous materials predominate over lattice oscillation below 1 K.
This model is referred to as the TLS model.
The degree of freedom concerning heat absorption becomes one when absorption by the TLS becomes dominant below 1 K.
That is why the temperature dependence of the heat capacity switches from cubic to linear.
The temperature dependence of the heat conductivity below 1 K is also successfully explained by the TLS model.
\citet{Agladze+1994} showed that the temperature dependence of the absorption coefficient in far-infrared wavebands measured for amorphous powder are well described by the TLS model in laboratory experiments.
They were the first to propose that the TLS may contribute to the observed features of interstellar dust.
\citet{Meny+2007} performed theoretical calculations of the frequency dependence of the absorption coefficient based on the TLS model.
\citet{Paradis+2011} compared their models with the observed spectra of diffuse interstellar dust from far infrared through submillimeter wavebands.
They showed that the TLS model succeeds in reproducing the observed features, including the inverse correlation of the spectral index with dust temperature.

\citet{Jones_2009} proposed the idea that the AME
might originate from the resonance emission due to radiative transition between the TLS of amorphous dust.
The fact that the effect of the TLS appears below 1 K, indicates that the energy splitting between the TLS is about $10^{-4}$ eV, which just coincides with the observed frequency of the AME.
Therefore, the resonance emission from amorphous dust is an attractive possibility for the origin of AME.
The negative correlation between the AME temperature and intensity is naturally explained by the amorphous model since the intensity of the resonance emission decreases as the dust temperature increases \citep{Meny+2007}.
They assumed that the peak value of the absorption cross-section of the resonance process of the TLS is the geometric cross-section.
It is well known, however, that the absorption cross-section of a small particle at microwave wavelengths is much smaller than the geometric cross-section (e.g., \cite{Draine_1984}).
It is likely that their model overestimates the TLS contribution.
It is also still unclear what kind of physical characteristics of the amorphous dust can be extracted from the observation of AME.
Because of the potential possibility of the amorphous origin of AME, studies of the thermal emission of amorphous dust relying on microscopic physical processes based on the TLS model are required.

In this paper, the intensity and polarization spectral energy distributions (SEDs) modeling from far infrared through microwave wavelengths were conducted based on the TLS model of amorphous dust.
By comparing the model with observations, we studied whether the amorphous dust model is able to explain the diagnostics of the entire frequency range spectrum; e.g., the emission peak in the far infrared, the spectral index in submillimeter wavebands, the bump in the emission of the AME, and
the low polarization fraction of the AME.
We showed what kind of physical characteristics are required for the amorphous dust in order to explain the observations.
We adopted two archetypical AME objects, the Perseus molecular cloud (MC) and W43 for our comparison with the observations.
Both objects have intensive data on the intensity and polarization spectrum over a wide number of frequency bands.

In section \ref{sec:TLS}, fundamental quantities of amorphous materials to describe their optical properties are summarized. 
Details of the basics of the standard TLS model are introduced in appendix \ref{sec:TLSmodel}.
In section \ref{sec:intensity}, we show how the SEDs of the thermal emission from amorphous silicate dust respond to the physical parameters of the TLS model and compare model SEDs with observational data.
Then we move on to the polarized emission in section \ref{sec:polarization}.
In section \ref{sec:property}, we examine the properties of the amorphous silicate dust. 
Limitation of the present model and possible improvements are discussed in section \ref{sec:discussion}.
Our conclusions and a summary are presented in section \ref{sec:conclusion}.

\section{Summary of fundamental quantities of amorphous materials to describe their optical properties}
\label{sec:TLS}

Optical properties of an amorphous material are determined by its electric susceptibility (see the details in sections \ref{sec:intensity} and \ref{sec:polarization}).
In this section, we summarize how the electric susceptibilities due to the TLS and disordered charged distribution (DCD) are related to the micro physics of each process, respectively.
The details of the standard TLS model are described in appendix \ref{sec:TLSmodel}.
The basic equations of the TLS model and the DCD model can apply to both amorphous silicate material and amorphous carbon material since the physical mechanism behind each model does not depend on the material composition.
The differences between the amorphous silicate material and the amorphous carbon material appear in the differences of values of physical variables.

\subsection{TLS model}
\begin{figure}
  \centering
  \includegraphics{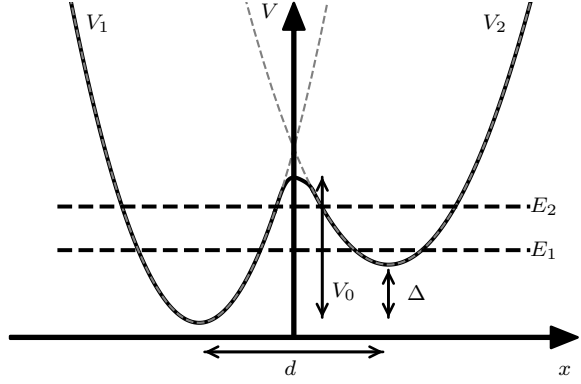}
  \caption{
  A double-well potential, shown by the black solid curve, in which an atom is trapped.
  The gray dashed curves denote harmonic potentials $V_1$ and $V_2$.
  }
  \label{fig:TLS_potential}
\end{figure}

The basic idea of the TLS model is that some of the atoms composing an amorphous material have two stable positions due to deformation of crystal structures.
The mechanical potential of the atom is described by the double-well potential illustrated in figure \ref{fig:TLS_potential}.
The $x$ coordinate marks the position of the atom.
This potential is generally described by a quartic function of $x$, which is called the soft-potential model (Karpov et al. \yearcite{Karpov+1982}).
In this paper, in order to describe how the TLS modifies the spectrum of the thermal emission from dust and determine whether AME can be explained by introducing the TLS model, we adopted the same approximation made by \citet{Anderson+1972} and \citet{Phillips_1972}, who first proposed the TLS model to describe the physical characteristics of the amorphous materials appearing at very low temperature.
They expanded the ground state and the first excited state of the Schr\"{o}dinger equation of the atom confined in the double-well potential $V$ by the two ground states when the atom is confined in each harmonic potential $V_1$ and $V_2$ individually (see figure \ref{fig:TLS_potential}).
We refer to this model as the standard TLS model.

As described in appendix \ref{sec:TLSmodel}, there are three independent processes, that is the resonance transition, the tunneling relaxation and the hopping relaxation, which contribute to the electric susceptibilities of the TLS. 
The complex susceptibilities\footnote{
  We define $\chi$ and $\chi_0$ as susceptibilities for the response to a macroscopic internal electric field and  to an external electric field, respectively. 
  $\chi$ is related to an electric permittivity $\varepsilon$ as 
  $\varepsilon = 1 + 4 \pi \chi$.
  }
for the resonance transition $\chi_0^\mathrm{res}$, the tunneling relaxation $\chi_0^\mathrm{tun}$ and the hopping relaxation $\chi_0^\mathrm{hop}$ are obtained as
\begin{eqnarray}
  \chi_0^\mathrm{res}
  &=&
  -i \int^{\Delta_0^\mathrm{max}}_{\Delta_0^\mathrm{min}} d\Delta_0
  \int^{\sqrt{(\Delta_0^\mathrm{max})^2 - \Delta_0^2}}_0 d\Delta
  f(\Delta_0,\ \Delta)
  \nonumber \\ && \times \
  \frac{\tau_+}{\hbar}
  \frac{| \bm{d}_0 |^2}{3}
  \left( \frac{\Delta_0}{E} \right)^2
  \tanh \left( \frac{E}{2 k_\mathrm{B} T} \right)
  \nonumber \\ && \times \
  \left[
    \frac{1}{1 + i(\omega - \omega_0) \tau_+}
  - \frac{1}{1 + i(\omega + \omega_0) \tau_+}
  \right]
  ,
  \label{eq:chi0_res}
  \\
  \chi_0^\mathrm{tun}
  &=&
  \int^{\Delta_0^\mathrm{max}}_{\Delta_0^\mathrm{min}} d\Delta_0
  \int^{\sqrt{(\Delta_0^\mathrm{max})^2 - \Delta_0^2}}_0
  d\Delta f(\Delta_0,\ \Delta)
  \nonumber \\ && \times \
  \frac{| \bm{d}_0 |^2}{3 k_\mathrm{B} T}
  \left( \frac{\Delta}{E} \right)^2
  \frac{1}{1 - i \omega \tau_\mathrm{tun}}
  \mathrm{sech} \left( \frac{E}{2 k_\mathrm{B} T} \right)
  ,
  \nonumber \\
  \label{eq:chi0_tun} \\
  \chi_0^\mathrm{hop}
  &=&
  \!\!\!\int^{\Delta_0^\mathrm{max}}_{\Delta_0^\mathrm{min}}\!\!\! d\Delta_0
  \!\!\!\int^{\sqrt{(\Delta_0^\mathrm{max})^2 - \Delta_0^2}}_0\!\!\! d\Delta
  f(\Delta_0,\ \Delta)
  \!\!\!\int^\infty_0\!\!\! dV_0 g(V_0)
  \nonumber \\ && \times \
  \frac{| \bm{d}_0 |^2}{3 k_\mathrm{B} T}
  \left( \frac{\Delta}{E} \right)^2
  \frac{1}{1 - i \omega \tau_\mathrm{hop}}
  \mathrm{sech} \left(\frac{E}{2 k_\mathrm{B} T} \right)
  ,
  \label{eq:chi0_hop}
\end{eqnarray}
where the definitions of parameters and distribution functions are as follows:
$\Delta$ is the energy difference between the two states located at each minimum of the double-well potential (see figure \ref{fig:TLS_potential});
$\Delta_0$ is the parameter that characterizes the degree of the cross correlation between the states located in two minima; 
$f(\Delta_0,\ \Delta)d\Delta_0d\Delta$ provides the number density of the atoms trapped in the TLS from $\Delta_0$ to $\Delta_0+d\Delta_0$ and from $\Delta$ to $\Delta+d\Delta$;
$\bm{d}_0$ is the electric dipole moment for the state located at the minimum of the potential $V_2$;
$E$ ($\equiv(\Delta^2+\Delta_0^2)^{1/2}$) is the energy splitting of the TLS;
$\tau_+$ is phase relaxation time;
$\tau_\mathrm{tun}$ is tunneling relaxation time;
$\tau_\mathrm{hop}$ is hopping relaxation time;
$\omega_0$ is defined as $E/\hbar$; 
$\omega$ is angular frequency of the electromagnetic wave that stimulates the resonance transition, the tunneling relaxation and the hopping relaxation;
$V_0$ is the height of the potential barrier;
$g(V_0)$ is probability density function. 
Upper and lower cutoff of $\Delta_0$, $\Delta_0^\mathrm{max}$ and $\Delta_0^\mathrm{min}$, are introduced to avoid divergence of the probability distribution function.
The typical values of physical variables of amorphous silicate material are given in table \ref{tab:fixed_param}.

\begin{table*}
  \tbl{Typical values of physical variables of an amorphous silicate material}{
  \begin{tabular}{llccc}
    \hline
    Parameter & Meaning & Value  & Unit & Refs. \footnotemark[$\ddagger$] \\
    \hline
    $V_m / k_\mathrm{B}$ &
    mean value of the barrier height distribution & 
    550 & K & 1 \\
    $V_\sigma / k_\mathrm{B}$ & 
    deviation of the barrier height distribution & 
    410 & K & 1 \\
    $V_\mathrm{min} / k_\mathrm{B}$ & 
    lower cutoff of the barrier height distribution &
    50 & K & 1 \\
    $\rho$ &
    mass density of a dust particle & 
    3.5 & g cm$^{-3}$ & 2 \\
    $n_\mathrm{atom}$ & 
    the number density of atoms composing a dust particle&
    $\rho /172.2 \times 7 N_\mathrm{A}$ \footnotemark[$*$] & cm$^{-3}$ & 2 \\
    $c_\mathrm{t}$ &
    sound velocity for transverse waves & 
    $3 \times 10^5$ & cm s$^{-1}$ & 1 \\
    $\gamma_\mathrm{t}$ &
    elastic dipole for transverse waves & 
    1 & eV & 3 \\
    $|\bm{d}_0|$ &
    electric dipole moment for the localized state at the bottom of $V_2$ & 
    1 & D & 1 \\
    $\tau_\mathrm{hop}^0$ & 
    pre-exponential factor for hopping relaxation time (see equation (\ref{eq:tau_hop}))& 
    10$^{-13}$ & s & 1 \\
    $l_\mathrm{c}$& 
    correlation length of propagation of lattice vibration & 
    30 & \AA & 1 \\
    $q^2$ &
    electric charge of an atom composing a dust particle& 
    $e^2$ & erg cm & 2 \\
    $m$ &
    mass of an atom composing a dust particle& 
    $m_\mathrm{O}$ \footnotemark[$\dagger$] & g & 2 \\
    $\omega_\mathrm{D}$ &
    Debye angular frequency &
    $2 \pi c_\mathrm{t} [9 n_\mathrm{atom} / (8 \pi)]^{1/3}$ & s$^{-1}$ & --- \\
    $C$ &
    correction factor for the DCD model &
    $4.15 \times 10^{-2} $ & --- & ---
    \\ \hline
  \end{tabular}}
  \label{tab:fixed_param}
  \begin{tabnote}
    \footnotemark[$*$]
    $N_\mathrm{A}$ is the Avogadro constant and we consider the composition of amorphous silicate materials as $\mathrm{MgFeSiO_4}$ 
    whose mass number is 172.2 g mol$^{-1}$. \\
    \footnotemark[$\dagger$]
    $m_\mathrm{O}$ is the mass of an oxygen atom.
     \\
     \footnotemark[$\ddagger$]
    References:
    1: \citet{Bosch_1978};
    2: \citet{Li+2001};
    3: \citet{Meny+2007}
  \end{tabnote}
\end{table*}

\subsection{Disordered charged distribution model}
Electric polarization due to the acoustic vibration propagating through the solid also makes a significant contribution to the absorption coefficient of the amorphous material.
To describe the irregular distribution of the lattice in the amorphous material, \citet{Schlomann_1964} introduced the DCD model.
The electric susceptibility derived from the DCD model, $\chi_0^\mathrm{DCD}$, is given as
\begin{eqnarray}
  \chi_0^\mathrm{DCD}
  &=&
  C
  \frac{q^2}{3 \pi^2 m}
  \int^{\omega_\mathrm{D} / c_\mathrm{t}}_0 dk
  \frac{k^2}{(c_\mathrm{t} k)^2 - \omega^2 + i \gamma \omega}
  h(k l_\mathrm{c}) ,
  \label{eq:chi0_DCD} \\
  h(x)
  &\equiv&
  1 - \frac{1}{(1 + x^2)^2} ,
\end{eqnarray}
where $l_\mathrm{c}$ is the correlation length of propagation of lattice vibration, $\omega_\mathrm{D}$ is the Debye frequency, $\gamma$ is the damping factor, and $C$ is the correction factor. 
The correction factor $C$ is introduced for the imaginary part of the dielectric susceptibility at 300 \micron \ predicted by DCD model so as to coincide with the 
imaginary part of the dielectric susceptibility at the same frequency proposed by \citet{Draine_1984}.  
The uncertainty of the adopted parameters listed in table \ref{tab:fixed_param} are absorbed by introducing the correction factor.
A crystalline material with a regular distribution is realized in the infinite correlation length limit, i.e., $k l_\mathrm{c} \gg 1$.
In this limit, the DCD model reduces to the Lorentz model.
In the limit of small $\gamma$, the following analytical formulae of $\chi_0^\mathrm{DCD}$ are obtained:
\begin{eqnarray}
  \mathrm{Re} \left( \chi_0^\mathrm{DCD} \right)
  =
  C
  \frac{q^2}{3 \pi^2 m c_\mathrm{t}^3}
  \Biggl[
  \omega_\mathrm{D}
  -\frac{\omega_\mathrm{c}^4 \omega_\mathrm{D}}
  {2 (\omega_\mathrm{c}^2 + \omega_\mathrm{D}^2)
  (\omega_\mathrm{c}^2 + \omega^2)}
  \nonumber \\
  +\frac{(2 \omega_\mathrm{c}^2 + \omega^2) \omega^3}
  {2 (\omega_\mathrm{c}^2 + \omega^2)^2}
  \ln
  \left| \frac{\omega_\mathrm{D} - \omega}{\omega_\mathrm{D} + \omega} \right|
  \nonumber \\
  -\frac{\omega_\mathrm{c}^3 (\omega_\mathrm{c}^2 - \omega^2)}
  {2 (\omega_\mathrm{c}^2 + \omega^2)^2}
  \mathrm{atan} \left( \frac{\omega_\mathrm{D}}{\omega_\mathrm{c}} \right)
  \Biggr] ,
  \label{eq:chi0_DCD_re}
  \\
  \left. \mathrm{Im} \left( \chi_0^\mathrm{DCD} \right)
  \right|_{\omega < \omega_{D}}
  =
  C
  \frac{q^2 \omega}{6 \pi m c_\mathrm{t}^3}
  \left\{1 - \frac{1}{\left[1 + (\omega / \omega_\mathrm{c})^2 \right]^2}
  \right\} ,
  \label{eq:chi0_DCD_im}
\end{eqnarray}
where $\omega_\mathrm{c} \equiv c_\mathrm{t}/l_\mathrm{c}$.
Although $\mathrm{Im}(\chi_0^\mathrm{DCD})$ becomes zero when $\omega > \omega_\mathrm{D}$,
it does not affect our analysis since we are interested in the low frequency range.

\section{Thermal emission from amorphous silicate dust}
\label{sec:intensity}
\subsection{Absorption cross section}

The absorption cross section of an amorphous dust is obtained by summing up contributions from the TLS and the DCD.
The electric susceptibility of the amorphous dust is written as
\begin{eqnarray}
  \chi_0
  &=&
  f_\mathrm{TLS}
  \left( \chi_0^\mathrm{res} + \chi_0^\mathrm{tun} + \chi_0^\mathrm{hop} \right)
  + \chi_0^\mathrm{DCD}
  ,
  \label{eq:chi0_amo}
\end{eqnarray}
where $f_\mathrm{TLS}$ is the fraction of atoms trapped in the TLS.

The absorption cross section is derived under the dipole approximation since the radius of the dust grain is much smaller than the wavelength, $\lambda$, of electromagnetic waves. 
In this section, the dust shape is assumed to be spherical.
The absorption cross section of the spherical amorphous dust, $C_\nu^\mathrm{abs}$, is given by the optical theorem as (c.f. \cite{Schlomann_1964}; \cite{Bohren+1983}; \cite{Meny+2007}),
\begin{eqnarray}
  C_\nu^\mathrm{abs}
  &=&
  \frac{8 \pi^2 \mathcal{V}}{\lambda}
  \mathrm{Im} \left(\chi_0 \right) 
  ,
  \label{eq:Cabs}
\end{eqnarray}
where $\mathcal{V}$ is the volume of an amorphous dust.

\subsection{Intensity emission spectra of thermal emission}

An intensity emission spectrum of the thermal emission of amorphous dust is deduced in this subsection.
In this paper, we take into account only an amorphous silicate dust and do not consider the contribution of carbonaceous dust such as PAHs, graphite, and amorphous carbon 
(expected effects of including amorphous carbon dust are discussed in section \ref{sec:discussion}).
The general formula for the spectrum of thermal emission from dust grains is given by
\begin{eqnarray}
  I_\nu^\mathrm{dust}
  = N_\mathrm{dust} \int da \frac{dn}{da} B_\nu(T) C_\nu^\mathrm{abs}(a,\ T)
  ,
  \label{eq:Inu_dust}
\end{eqnarray}
where $N_\mathrm{dust}$ is the column density of the dust grains in the line of sight, and $dn/da$ provides the size distribution of the dust grains and is normalized to be 1 by integrating over the dust size $a$.
The size distribution function proposed by \citet{Draine+2007} is adopted.
In this study, we neglect the effect of the time variation in the temperature of small dust grain to the SED of the thermal emission of amorphous silicate dust. 
We assume that all dust grains stay at the same temperature.

\begin{figure*}[!t]
  \begin{minipage}{0.5\hsize}
    \centering
    \subcaption{$\Delta_0^\mathrm{max}$}
    \includegraphics{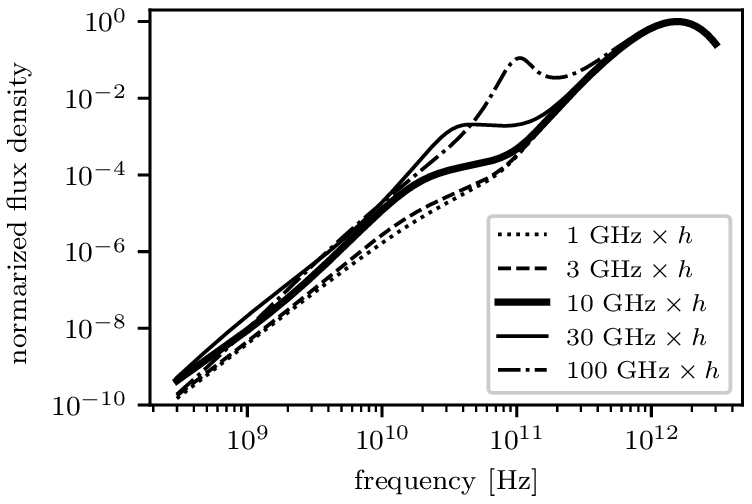}
    \label{fig:SED_D0max}
  \end{minipage}
  \begin{minipage}{0.5\hsize}
    \centering
    \subcaption{$R_\Delta$}
    \includegraphics{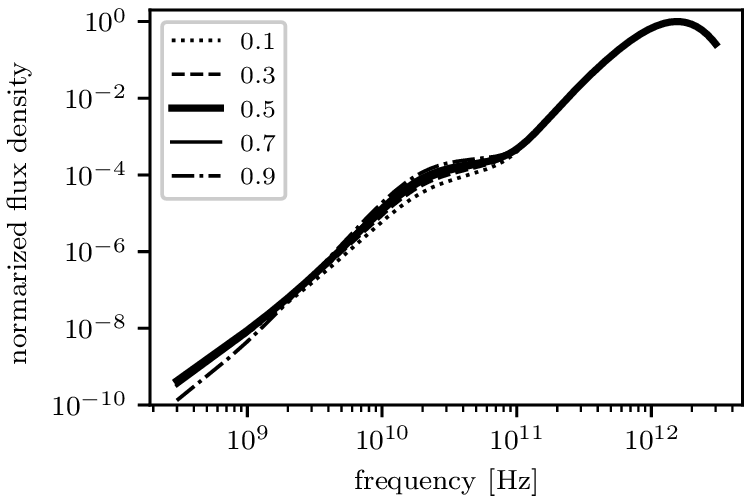}
    \label{fig:SED_D0min}
  \end{minipage}
  \begin{minipage}{0.5\hsize}
    \centering
    \subcaption{$T$}
    \includegraphics{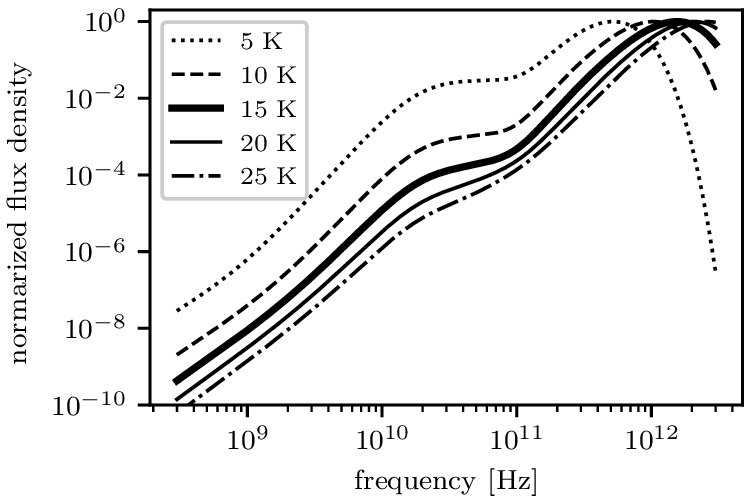}
    \label{fig:SED_temp}
  \end{minipage}
  \begin{minipage}{0.5\hsize}
    \centering
    \subcaption{$\tau_+$}
    \includegraphics{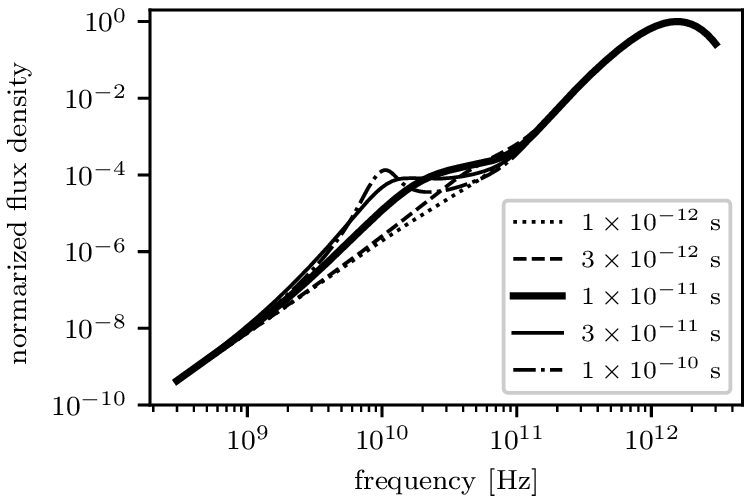}
    \label{fig:SED_tau}
  \end{minipage}
  \begin{minipage}{1\hsize}
    \centering
    \subcaption{$f_\mathrm{TLS}$}
    \includegraphics{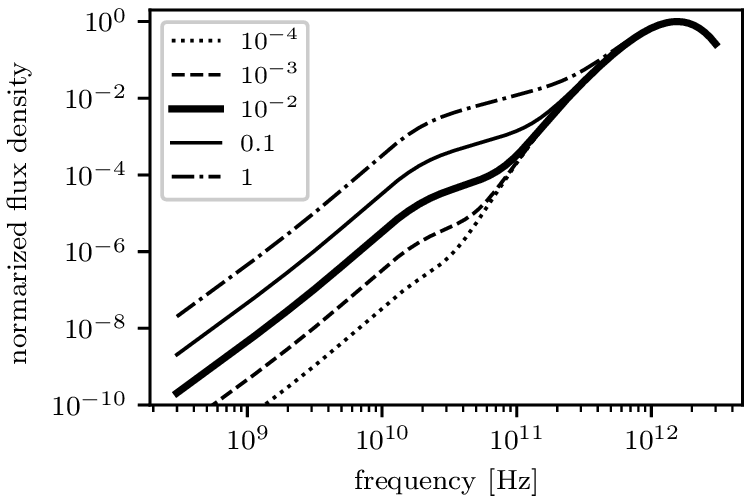}
    \label{fig:SED_fTLS}
  \end{minipage}
  \caption{
  Parameter dependences of the SEDs of dust thermal emission in the standard TLS model.
  SEDs are given in arbitrary units normalized to each maximum value.
  Thick solid curves in each panel show the SEDs with the same parameter values.
  In each panel, one of the variables characterizing the amorphous silicate dust was varied to see how the shape of the SED responds for each variable.
  The variables are (a) $\Delta_0^\mathrm{max}$ which are expressed in the corresponding frequency normalized by the Planck constant $h$,
  (b) $R_{\Delta}$, (c) $T$, (d) $\tau_+$, and (e) $f_\mathrm{TLS}$, respectively.
  Given values for each parameter are shown in legends. 
  }
  \label{fig:SED_TLS}
\end{figure*}

Figure \ref{fig:SED_TLS} shows the parameter dependence of thermal emission SEDs from amorphous silicate dust.
The results show that the bump emission appears at around several tens of GHz.
These are caused by the resonance transition of the TLS.
Figure \ref{fig:SED_D0max} shows that the peak frequency of the bump emission is shifted toward higher frequency as the upper limit of the energy difference between the TLS, $\Delta_0^\mathrm{max}$, increases while $R_\Delta$ ($\equiv\Delta_0^\mathrm{min}/\Delta_0^\mathrm{max}$) is fixed.
Figure \ref{fig:SED_D0min} shows that the bump feature of the resonance emission becomes broader as $R_\Delta$ gets smaller, although the response is not prominent.
Figure \ref{fig:SED_temp} shows that the bump emission due to the resonance process relative to the far-infrared peak becomes higher when the temperature of the dust grain lowers.
This is attributed to the fact that, the electric dipole moment caused by the resonance transition rate increases with decreasing temperature because the fraction of atoms in the ground state increases with decreasing temperature (see appendix \ref{sec:TLSmodel} equation (\ref{eq:dipole_res})).
Figure \ref{fig:SED_tau} shows that the width of the bump emission sensitively responds to the relaxation time scale of the resonance process, $\tau_+$.
Figures \ref{fig:SED_D0max} and \ref{fig:SED_tau} show that the bump emission becomes prominent when $1/\tau_+$ becomes comparable to, or greater than, $\Delta_0^\mathrm{max} / h$.
Figure \ref{fig:SED_fTLS} shows that the peak intensity of the bump emission caused by the resonance process is about two orders of magnitude lower than the peak intensity in the far infrared, even when all the atoms are trapped in the TLS.
The relative intensity of the bump emission decreases almost linearly with $f_\mathrm{TLS}$.

\begin{figure}[t]
  \centering
  \includegraphics{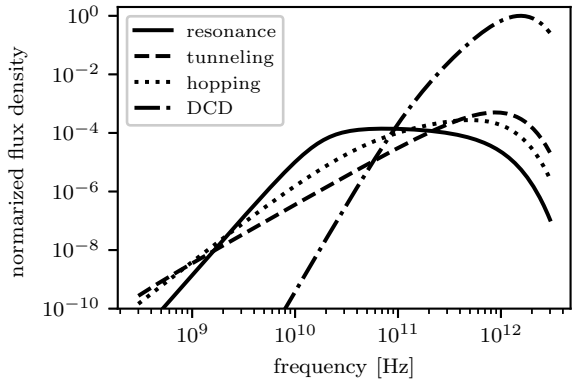}
  \caption{
  Dust thermal emission SEDs caused by each TLS and DCD process.
  Parameters have the same values as the thick solid curves in figure \ref{fig:SED_TLS}.
  Every curve is normalized to the peak value of the SED arising from the DCD.
  }
  \label{fig:SED_sep}
\end{figure}
To clarify how the frequency dependence of the thermal emission of amorphous silicate dust is defined, figure \ref{fig:SED_sep} shows SEDs for each process.
The frequency dependence of the absorption coefficient of the resonance process and the relaxation processes in submillimeter wavebands are described by $C_\nu^\mathrm{res} \propto \nu^2$ and $C_\nu^\mathrm{rel} \propto \nu$, respectively.
The frequency dependence of the resonance process in the long wavelength limit is the same as for crystal.
As the wavelength increases starting from the far infrared, the contributions from tunneling and hopping relaxation become more significant.
As a result, the slope of the absorption coefficient of the amorphous material becomes flatter than that of crystal in the submillimeter wavelength range.

\subsection{Comparison with observed spectra}
\subsubsection{Intensity SED data and modeling}

Our model SEDs are fitted to the observed spectra from millimeter through far infrared for two MCs, Perseus and W43, for which prominent AME is detected. 
The observed data for the Perseus MC and W43 are taken from table 2 in \citet{QUIJOTE1_2015} and table 3 in \citet{QUIJOTE2_2017}, respectively.
Although the temperature fluctuation of the CMB is subtracted from the spectrum of W43, it is not taken into account in the spectrum of the Perseus MC.
Therefore, the SED fit with and without the CMB contribution are performed for the Perseus MC.
Planck data at 100 and 217 GHz may still be contaminated by CO residuals \citep{PlanckXX_2011}.
To take this possibility into account, data points at 100 and 217 GHz are not included in the fit for the Perseus MC with a CMB contribution.
These frequency bands are included in the fit for other cases. 
The observed spectra of these MCs are shown in figure \ref{fig:fit_SED}.
The contributions of synchrotron emission, free--free emission and dust thermal emission from the Galactic interstellar medium along the line of sight were removed by subtracting the median value of the intensity surrounding each MC.
As for the SED of the free--free emission originating from each MC, the formulae adopted by \citet{PlanckXX_2011} are applied in this paper; that is,
\begin{eqnarray}
  I_\nu^\mathrm{ff}
  &=& \frac{2 k_\mathrm{B} T_\mathrm{ff} \nu^2}{c^2} 
  ,
  \label{eq:Inu_ff} \\
  T_\mathrm{ff}
  &=& T_\mathrm{e} (1 - e^{-\tau_\mathrm{ff}}) 
  ,
  \label{eq:Tff} \\
  \tau_\mathrm{ff}
  &=&
  3.014 \times 10^{-2} T_\mathrm{e}^{-1.5}
  \left(\frac{\nu}{\mathrm{GHz}} \right)^{-2}
  \left(\frac{\mathrm{EM}}{\mathrm{cm^{-6}\ pc}} \right) g_\mathrm{ff} 
  ,
  \label{eq:tauff} \\
  g_\mathrm{ff}
  &=&
  \ln \left[ 4.955 \times 10^{-2}
  \left( \frac{\nu}{\mathrm{GHz}} \right)^{-1} \right]
  + 1.5 \ln \left(\frac{T_\mathrm{e}}{\mathrm{K}} \right) 
  ,
  \label{eq:gff}
\end{eqnarray}
where $\mathrm{EM}$ is the emission measure.
The electron temperature of each MC is fixed at $T_\mathrm{e} = 8000$ K for the Perseus MC \citep{PlanckXX_2011} and $T_\mathrm{e} = 6038$ K for W43 \citep{Alves+2012}.
Therefore, free parameters to fit the observed SEDs are EM which characterizes the fraction of the free--free contribution, dust temperature $T$, dust column density $N_\mathrm{dust}$, the fraction of the atoms trapped in the TLS $f_\mathrm{TLS}$, the upper and lower bounds of $\Delta_0$ (that is, $\Delta_0^\mathrm{max}$ and $\Delta_0^\mathrm{min}$), and the relaxation time scale $\tau_+$.
In the case of the Perseus MC, the amplitude of the temperature fluctuation of the CMB $\Delta T_\mathrm{CMB}$ is also an additional fit parameter.
The SED of the CMB temperature fluctuation is given as, 
\begin{eqnarray}
  I_\nu^{\mathrm{CMB}}
  &=&
  B_\nu (T_\mathrm{CMB}) \frac{x e^x}{e^x - 1} 
  \frac{\Delta T_\mathrm{CMB}}{T_\mathrm{CMB}}
  ,
  \label{eq:Inu_cmb}
  \\
  x &=& \frac{h \nu}{k_\mathrm{B} T_\mathrm{CMB}}
  ,
\end{eqnarray}
where $T_\mathrm{CMB} = 2.725 \ \mathrm{K}$ \citep{Mather+1999} is the CMB temperature.

\subsubsection{Fit results}
\label{sec:intensity_result}

\begin{figure*}[t]
  \begin{minipage}[t]{0.5\hsize}
    \centering
    \subcaption{Perseus MC}
    \includegraphics{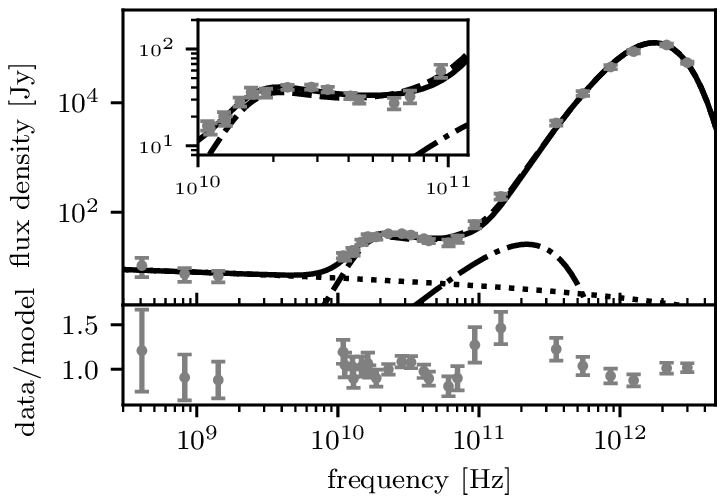}
    \label{fig:fit_per}
  \end{minipage}
  \begin{minipage}[t]{0.5\hsize}
    \centering
    \subcaption{W43}
    \includegraphics{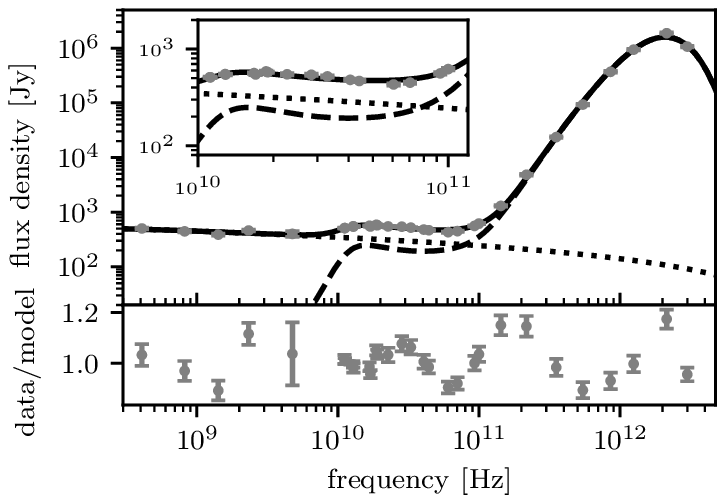}
    \label{fig:fit_w43}
  \end{minipage}
  \caption{
  Spectra of two archetypal dust-rich objects accompanying prominent AME, (a) the Perseus MC and (b) W43, are fitted by our amorphous model.
  Dashed curves are the best fit thermal emission model from amorphous silicate dust, and dotted lines are the best fit free--free emission contribution.
  In both data, contributions from the Galactic interstellar medium are subtracted.
  Although temperature fluctuation of the CMB is subtracted from the spectrum of W43, it is not taken into account in the spectrum of the Perseus MC.
  Therefore, the CMB contribution is taken into account in the SED fit for the Perseus MC. 
  As shown in table \ref{tab:fit_param}, the best CMB temperature fluctuation takes negative value.
  The absolute value of the best-fit CMB contribution is shown by dashed dotted line in (a).
  The solid curves show the total SEDs of the best-fit models. 
  The bottom panel inserted in each figure shows the data-to-model ratio. 
  }
  \label{fig:fit_SED}
\end{figure*}
We searched the parameters that minimize the chi squared by a brute force.
The best-fitting model SEDs based on our amorphous model are overlaid on the observed spectra in figure \ref{fig:fit_SED}.
As shown in table \ref{tab:fit_param}, the best CMB temperature fluctuation takes a negative value.
The absolute value of the best-fit CMB contribution is shown by dashed-dotted line in figure \ref{fig:fit_per}.
The best-fit parameters are summarized in table \ref{tab:fit_param}.
\begin{table}[t]
  \tbl{Best-fit parameters for the Perseus MC and W43}{
  \begin{tabular}{lr@{$\pm$}lr@{$\pm$}lr@{$\pm$}l}
    \hline
    \multicolumn{1}{l}{} & \multicolumn{4}{c}{Perseus MC} & \multicolumn{2}{c}{W43}
    \\
    \multicolumn{1}{l}{} & \multicolumn{2}{c}{without} & 
    \multicolumn{2}{c}{with CMB} & \multicolumn{2}{l}{} \\
    \hline
    $T \ (\mathrm{K})$ &
    16.78 & 0.08 & 
    16.90 & 0.08 &
    20.20 & 0.06
    \\
    $\tau_{250}\ (\times 10^{-4})$ &
    4.08 & 0.08 & 
    3.87 & 0.07 & 
    58.3 & 0.5
    \\
    $f_\mathrm{TLS}$ &
    0.0123 & 0.0003 & 
    0.0151 & 0.0004 &
    0.0338 & 0.0004
    \\
    $\Delta_0^\mathrm{max}/h \ (\mathrm{GHz})$ &
    15.2 & 0.2 & 
    15.0 & 0.2 & 
    11.0 & 0.1
    \\
    $R_\Delta$ &
    0.716 & 0.023 & 
    0.648 & 0.022 & 
    0.927 & 0.006
    \\
    $\tau_+ \ (\times 10^{-11} \ \mathrm{s})$ &
    2.24 & 0.09 & 
    \multicolumn{2}{c}{$2.19^{+0.09}_{-0.08}$} & 
    2.89 & 0.05
    \\
    EM (cm$^{-6}$ pc) &
    26.9 & 2.4 & 
    26.7 & 2.4 & 
    3934 & 29
    \\
    \multicolumn{1}{l}{$\Delta T_\mathrm{CMB} \ (\mu \mathrm{K})$} &
    \multicolumn{2}{c}{---} &
    \multicolumn{2}{c}{$-19.3^{+6.3}_{-5.9}$} &
    \multicolumn{2}{c}{---}
    \\
    \multicolumn{1}{l}{dof} &
    \multicolumn{2}{c}{21} &
    \multicolumn{2}{c}{18} &
    \multicolumn{2}{c}{23}
    \\
    \multicolumn{1}{l}{$\chi^2/\mathrm{dof}$} &
    \multicolumn{2}{c}{2.04} &
    \multicolumn{2}{c}{1.67} &
    \multicolumn{2}{c}{6.41}
    \\ \hline
  \end{tabular}} \label{tab:fit_param}
  \begin{tabnote}
    The errors are at $1\sigma$ errors.
  \end{tabnote}
\end{table}
Assuming an optically thin condition, $N_\mathrm{dust}$ is interpreted as the optical depth at $\lambda = 250 \ \micron$, $\tau_{250}$.
Our SED models reproduce the observed SEDs from AME through the far infrared feature very well.
In our models, AME originates mainly from the resonance emission of the TLS of large grains.
It should be stressed that the amorphous model is able to explain AME without introducing new species.
The reduced chi squared of the best-fit models for the Perseus MC is $\chi^2 / \mathrm{dof} = 2.04$, where $\mathrm{dof} = 21$ without CMB, $\chi^2 / \mathrm{dof} = 1.67$, where $\mathrm{dof} = 18$ with CMB, and for W43 $\chi^2 / \mathrm{dof} = 6.41$ where $\mathrm{dof} = 23$.
For W43, the overall observed feature is also well reproduced by our model although the quality of the fit is not so good.
The bottom panels of figures \ref{fig:fit_per} and \ref{fig:fit_w43} show that our models underestimate the observed intensities in the frequency range from 100 GHz through 500 GHz.

\section{Polarized emission}
\label{sec:polarization}

The observations of polarization emission is one of the crucial keys to discriminating among the models of the origin of AME.
Dust thermal emission is supposed to be polarized because the shapes of the dust grains are non-spherical and align with the magnetic field.
Hereafter, dust shape is represented by an ellipsoid for simplicity.
In this section, the theoretical model of polarized emission from amorphous silicate dust based on the standard TLS model is established and the model predictions are compared with the observed results obtained for the Perseus MC and W43.

\subsection{Absorption and polarization cross section for ellipsoidal dust}
\label{sec:ellipsoid}

The shape of an ellipsoid is characterized by the radii of three axes, a radius of semi-major axis $a_x$, semi-middle axis $a_y$ and semi-minor axis $a_z$, that is $a_x \ge a_y \ge a_z$.
We take the semi-major axis along the $x$-axis, the semi-middle axis along the $y$-axis, and the semi-minor axis along the $z$-axis.

\begin{figure}
  \centering
  \includegraphics{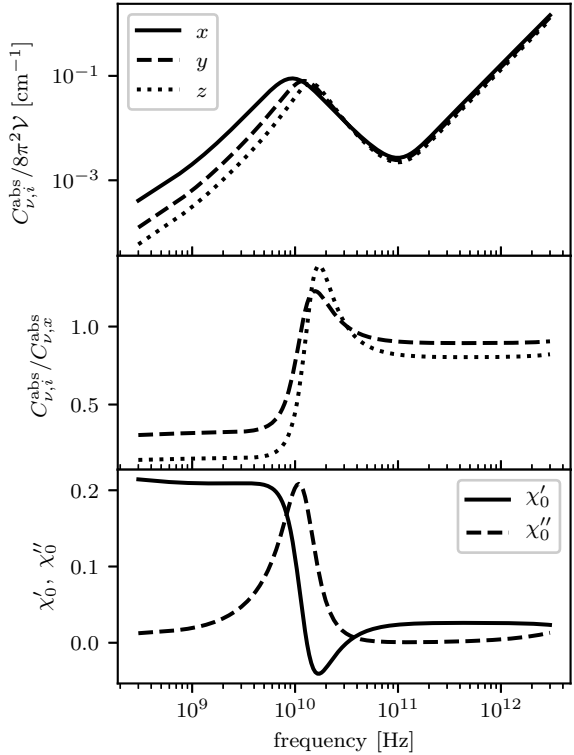}
  \caption{
  The frequency dependence of absorption cross sections for an electric field parallel to each axis of an ellipsoid normalized by $8 \pi^2 \mathcal{V}$ (top panel), of relative absorption cross sections for electric field parallel to semi-middle and semi-minor axis to that of the semi-major axis (middle panel), and of the real and imaginary parts of the electric susceptibility for a spherical dust (bottom panel). 
  A ratio of geometric factors is fixed to  $L_x : L_y : L_z = 1 : 2 : 3$. 
  Other parameters are set to the best-fit values for W43 listed in table \ref{tab:fit_param}.
  }
  \label{fig:Cabs_elip}
\end{figure}
The absorption cross sections of an ellipsoidal particle for radiation linearly polarized along each axis are given by following equation:
\begin{eqnarray}
  C_{\nu,i}^\mathrm{abs}
  =
  \frac{8\pi^2\mathcal{V}}{\lambda}
  \mathrm{Im} \left(\chi_0^i\right)
  \label{eq:Cabs_i_main}
  ,
\end{eqnarray}
where $i=x,\ y,\ z$ and $\chi_0^i$ are the complex susceptibilities responding to an external electric field parallel to each axis. 
As shown in appendix \ref{sec:localF}, $\chi_0^i$ are given by, 
\begin{eqnarray}
  \chi_0^i
  =
  \frac{\chi_0}{1+4\pi(L_i-1/3)\chi_0}
  ,
  \label{eq:chi0_relation_main}
\end{eqnarray}
where $L_i$ are geometric factors defined as \citep{Bohren+1983},
\begin{eqnarray}
  L_i
  \equiv
  \frac{3\mathcal{V}}{8\pi}
  \int^\infty_0\frac{dq}{(q+a_i^2)\sqrt{(q+a_x^2)(q+a_y^2)(q+a_z^2)}}
  .
  \label{eq:Li_msin}
\end{eqnarray}
Figure \ref{fig:Cabs_elip} shows the frequency dependences of $C_{\nu,x}^\mathrm{abs}$, $C_{\nu,y}^\mathrm{abs}$, and $C_{\nu,z}^\mathrm{abs}$ for amorphous silicate dust.
The adopted values of the geometrical factors are $L_x = 1/6$, $L_y = 1/3$, and $L_z = 1/2$. 
In general, $C_{\nu, x}^\mathrm{abs}$ takes the largest value and $C_{\nu, y}^\mathrm{abs}$ takes the median value of the three. 
This can be understood by a change of sign of the term $4 \pi (L_i - 1/3) \chi_0$ appearing in the denominator  of equation (\ref{eq:chi0_relation_main}). 
For the semi-major axis, this term takes a negative sign. 
On the other hand, this term is zero for the semi-middle axis and is positive for the semi-minor axis. 
Therefore, the denominator of equation (\ref{eq:chi0_relation_main}) takes the smallest value for the semi-major axis and the largest value for the semi-minor axis. 
The above-mentioned order of the amplitude of the absorption cross section is a consequence of this result.  
However, the order of the amplitude becomes reversed at around the resonance peak. 
This is evident in the middle panel of figure \ref{fig:Cabs_elip}. 
Figure \ref{fig:Cabs_elip} shows that the resonant peak frequency for the semi-middle axis coincides with that of the imaginary part of the electric susceptibility for the spherical particle. 
This is the expected result since $L_y=1/3$. 
The resonant peak for the semi-major axis appears at a slightly lower frequency than the peak frequency of the imaginary part of the electric susceptibility for the spherical particle. 
This reflects the fact that $L_x$ is smaller than $1/3$. 
The resonant peak frequency of the absorption cross section for the semi-minor axis is shifted to a higher frequency since $L_z>1/3$.

In order to model the polarization emission from amorphous silicate dust we make following simplifications.
The semi-minor axis of each dust grain is perfectly aligned with the magnetic field, and the directions of the semi-major axis around the magnetic field are randomly distributed.
The magnetic field is uniformly distributed along the line of sight and the direction of the field is perpendicular to the line of sight.
These simplifications maximize the prediction of the polarization degree based on our models.
The ensemble average of the absorption and polarization cross sections $\langle C_\nu^\mathrm{abs} \rangle$ and  $\langle C_\nu^\mathrm{pol} \rangle$ are deduced by averaging over the direction of the semi-major axis around the magnetic field, as in \citet{Draine+2017}:
\begin{eqnarray}
  \langle C_\nu^\mathrm{abs} \rangle
  &=&
  \frac{
  \langle C_{\nu,x}^\mathrm{abs} \rangle 
  + \langle C_{\nu,y}^\mathrm{abs} \rangle 
  + 2 \langle C_{\nu,z}^\mathrm{abs} \rangle }
  {4}
  ,
  \label{eq:Cabs_CDE} \\
  \langle C_\nu^\mathrm{pol} \rangle
  &=&
  \frac{
  \langle C_{\nu,x}^\mathrm{abs} \rangle
  + \langle C_{\nu,y}^\mathrm{abs} \rangle
  - 2 \langle C_{\nu,z}^\mathrm{abs} \rangle}
  {4}
  ,
  \label{eq:Cpol_CDE}
\end{eqnarray}
where $\langle C_{\nu,x}^\mathrm{abs} \rangle$, $\langle C_{\nu,y}^\mathrm{abs} \rangle$ and $\langle C_{\nu,z}^\mathrm{abs} \rangle$ are the shape-averaged absorption cross sections for the linearly polarized radiation in the direction of each axis as defined in appendix \ref{sec:ellip}.
Except around the resonance peak frequency, 
$\langle C_\nu^\mathrm{pol} \rangle$ takes positive value 
since 
$\langle C_{\nu,x}^\mathrm{abs} \rangle 
> \langle C_{\nu,y}^\mathrm{abs} \rangle 
> \langle C_{\nu,z}^\mathrm{abs} \rangle$. 
Therefore, the predicted direction of the polarization emission is perpendicular to the magnetic field.

\begin{figure*}[tp]
  \begin{minipage}{0.5\hsize}
    \centering
    \subcaption{$\Delta_0^\mathrm{max}$}
    \includegraphics{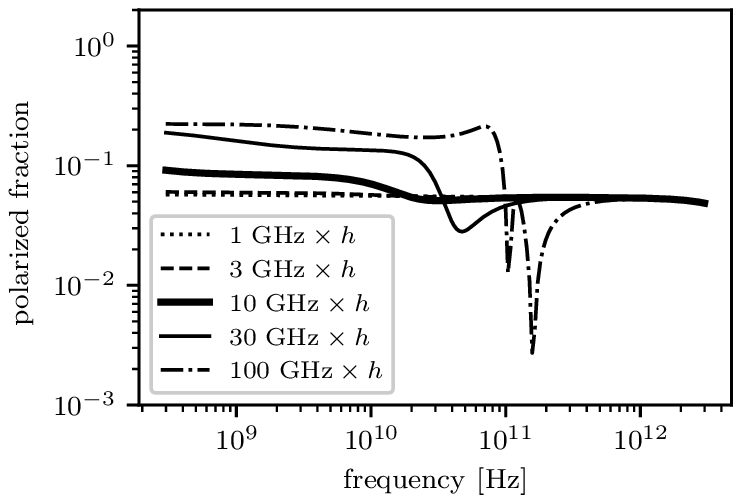}
    \label{fig:DoP_D0max}
  \end{minipage}
  \begin{minipage}{0.5\hsize}
    \centering
    \subcaption{$R_\Delta$}
    \includegraphics{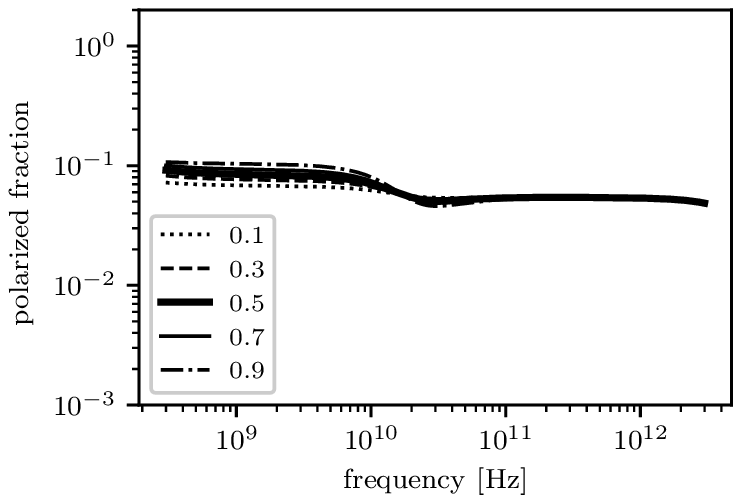}
    \label{fig:DoP_D0min}
  \end{minipage}
  \begin{minipage}{0.5\hsize}
    \centering
    \subcaption{$T$}
    \includegraphics{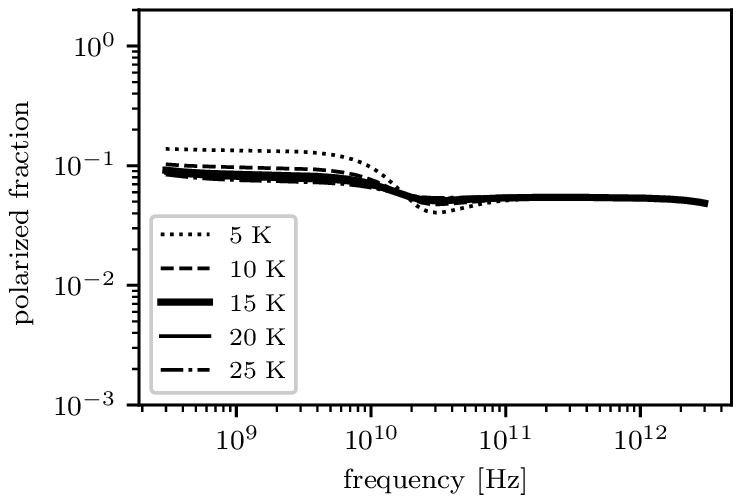}
    \label{fig:DoP_temp}
  \end{minipage}
  \begin{minipage}{0.5\hsize}
    \centering
    \subcaption{$\tau_+$}
    \includegraphics{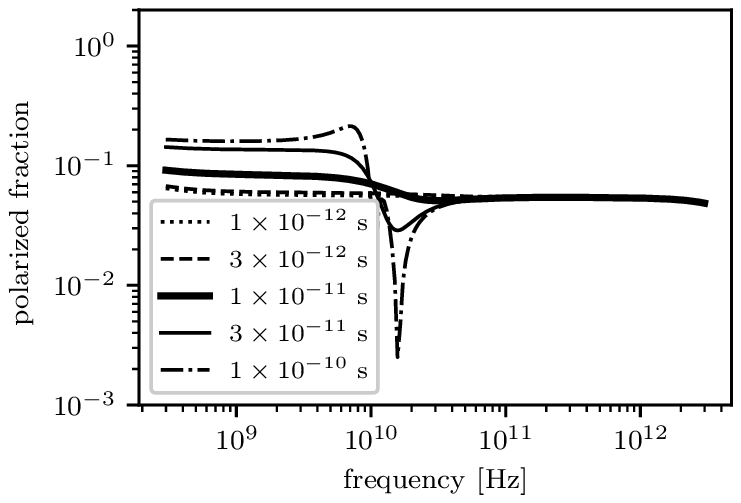}
    \label{fig:DoP_tau}
  \end{minipage}
  \begin{minipage}{1\hsize}
    \centering
    \subcaption{$L_\mathrm{min}$}
    \includegraphics{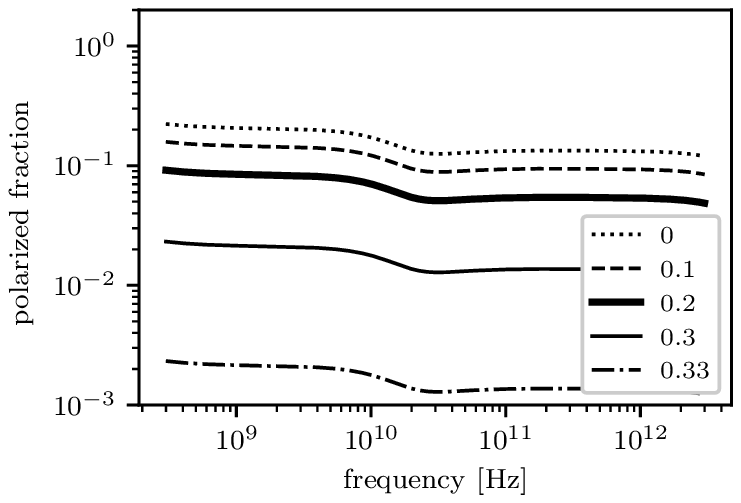}
    \label{fig:DoP_Lmin}
  \end{minipage}
  \caption{
  Parameter dependences of the degree of polarization of the thermal emission from amorphous silicate dust as a function of frequency. 
  For reference, the polarization degree with the same parameter set is shown by a solid line in each panel.
  The fraction of the atoms trapped in the TLS is set to be 0.01.
  In each panel, one of the variables characterizing the amorphous silicate dust is varied with respect to the reference model to see how the shape of the polarization degree responds to each variable.
  In the case of $\Delta_0^\mathrm{max}/h = 100$ GHz shown in (a), $\langle C_\nu^\mathrm{pol}\rangle$ defined by equation (\ref{eq:Cpol_CDE}) takes  negative values between 100 GHz and 500 GHz.  
  The polarization cross sections take positive values for all other cases shown in these figures. 
 The variables are (a) $\Delta_0^\mathrm{max}$, which are expressed in the corresponding frequency normalized by $h$, (b) $R_{\Delta}$, (c) $T$, (d) $\tau_+$, and (e) $L_\mathrm{min}$, respectively.
  The given values for each parameter are shown in the legends.
  }
  \label{fig:DoP_TLS}
\end{figure*}

The degree of polarization $\Pi_\nu$ is obtained by taking the ratio of $|\langle C_\nu^\mathrm{pol} \rangle|$ to $|\langle C_\nu^\mathrm{abs} \rangle|$.
Figure \ref{fig:DoP_TLS} shows the frequency dependence of $\Pi_\nu$ of the thermal emission from amorphous silicate dust.
In the frequency range, except in the waveband around the resonance peak, $\Pi_\nu$ is nearly constant.
Since $\langle C_\nu^\mathrm{pol}\rangle$ takes a positive value, the direction of polarization is perpendicular to the magnetic field,  as expected. 
Since the imaginary part of the susceptibility, $\chi_0''$, is much smaller than the real part, $\chi_0'$, in these frequency ranges, the ensemble averages of the absorption and polarization cross sections are expressed as
$\langle C_\nu^\mathrm{abs} \rangle = \chi_0'' f(\chi_0')$ and
$\langle C_\nu^\mathrm{pol} \rangle = \chi_0'' g(\chi_0')$ 
in the first-order of the imaginary part. 
Therefore, $\Pi_\nu$ is independent of $\chi_0''$ and depends only on $\chi_0'$.
Since the frequency dependence of $\chi_0'$ is very small, $\Pi_\nu$ of the high and low frequency ranges, except around the resonance peak, become almost constant against frequency change.
In the high frequency range, the DCD contribution of $\chi_0'$ is dominant.
On the other hand, in the low frequency part, the contribution of the resonance process to $\chi_0'$ is dominant.
This results in the discrepancy of $\Pi_\nu$ found in figure \ref{fig:DoP_TLS} between the high and low frequency region across the resonance peak.

At around the resonance peak, the degree of polarization shows a prominent behavior for some sets of parameters. 
Figures \ref{fig:DoP_D0max} and \ref{fig:DoP_tau} show that the degree of polarization decreases abruptly and takes the local minima around the peak frequency of the resonance process when $\Delta_0^\mathrm{max} > h/\tau_+$. 
This is because the amplitude of the polarization cross sections for all three axes of the ellipsoid get closer near the resonance peak, as shown in figure \ref{fig:Cabs_elip}.
As a result, the polarization cross section defined by equation (\ref{eq:Cpol_CDE}) approaches zero.  
In extreme cases, the polarization cross section changes its sign. 
This can be seen in the case of $\Delta_0^\mathrm{max}/h = 100$ GHz in figure \ref{fig:DoP_D0max}. 
In this case, the order of the amplitude of the absorption cross section reverses:
$\langle C_{\nu,x}^\mathrm{abs} \rangle 
< \langle C_{\nu,y}^\mathrm{abs} \rangle 
< \langle C_{\nu,z}^\mathrm{abs} \rangle$.
As a result, 
$\langle C_\nu^\mathrm{pol} \rangle$ becomes negative. 
This means that the direction of the polarization changes and becomes parallel to the magnetic field near the resonance peak frequency.
Figure \ref{fig:DoP_temp} shows that the polarization degree takes a minimum value at the resonance peak when the temperature of the dust is as low as 10 K. 
This is because the relative intensity of the resonance peak to far infrared emission increases when the dust temperature decreases, as shown in figure \ref{fig:SED_temp}.

\subsection{Comparison with astronomical data}

There is no definite report on the detection of the polarization from AME.
The upper limits on $\Pi_\nu$ for the Perseus MC and W43 are given by \citet{QUIJOTE1_2015} and \citet{QUIJOTE2_2017}, respectively.
In this subsection, we attempt to fit intensity and polarization data simultaneously with our model.

\subsubsection{Polarization SED data and modeling}

\begin{figure*}[t]
  \begin{minipage}[t]{0.5\hsize}
    \centering
    \subcaption{Perseus MC}
    \includegraphics{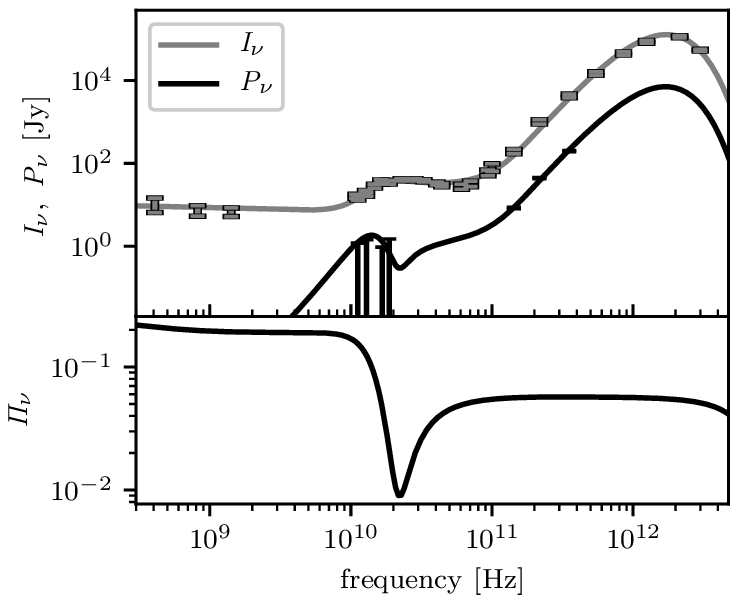}
    \label{fig:Pol_per}
  \end{minipage}
  \begin{minipage}[t]{0.5\hsize}
    \centering
    \subcaption{W43}
    \includegraphics{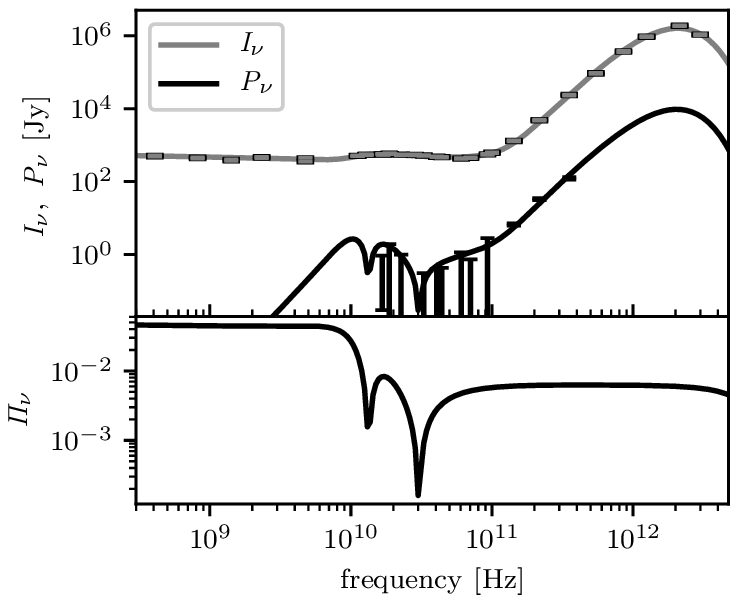}
    \label{fig:Pol_w43}
  \end{minipage}
  \caption{
  Intensity and polarized SEDs, and the polarization fraction for (a) the Perseus MC and (b) W43.
  The error bars are 1$\sigma$ but the upper limits are at the 95\% confidence level.
  }
  \label{fig:Pol_SED}
\end{figure*}

In figure \ref{fig:Pol_SED}, the polarized SEDs for the Perseus MC and W43 are shown.
The data for polarized AME are taken from table 4 in \citet{QUIJOTE1_2015} and table 8 in \citet{QUIJOTE2_2017}. 
We exclude the DRAO 1.4 GHz data point because \citet{QUIJOTE2_2017} suspected that Faraday rotation affects the data point. 
The polarization flux at 143, 217, and 353 GHz are extracted from the Planck second data release \citep{PlanckI_2016} using the same method adopted in \citet{QUIJOTE2_2017}. 
Although the data point at 23 GHz quoted from WMAP indicates the detection of polarization, we cannot reject the possibility that the subtraction of the Galactic synchrotron contribution might be insufficient, and that the contribution of the Galactic synchrotron is dominant in the data point \citep{QUIJOTE2_2017}.
Therefore, we treat this point as the upper limit when the fit is performed. 
The central values of the data points where the upper limits are given are set to zero.

The intensity and polarization SEDs of ellipsoidal amorphous silicate dust are given by substituting equations (\ref{eq:Cabs_CDE}) and (\ref{eq:Cpol_CDE}) for equation (\ref{eq:Inu_dust}). 
We include $L_\mathrm{min}$ among the fit parameters.
The free--free emission is assumed to be unpolarized.

\subsubsection{Fit results}

The observed SEDs of the intensity and polarization flux are fitted simultaneously. 
The brute-force fitting method adopted in subsection \ref{sec:intensity_result} is used.
The best-fit parameters are summarized in table \ref{tab:fit_param_pol}.
\begin{table}
  \tbl{best-fit parameters for the Perseus MC and W43 with polarization}{
  \begin{tabular}{lr@{$\pm$}lr@{$\pm$}l}
    \hline
    & \multicolumn{2}{c}{Perseus MC} & \multicolumn{2}{c}{W43} \\
    \hline 
    $T \ (\mathrm{K})$ &
    16.67 & 0.07 &
    20.14 & 0.06
    \\
    $\tau_{250}\ (\times 10^{-4})$ & 
    4.25 & 0.06 &
    59.3 & 0.5
    \\
    $f_\mathrm{TLS}$ & 
    0.0136 & 0.0004 &
    0.0293 & 0.04
    \\
    $\Delta_0^\mathrm{max}/h \ (\mathrm{GHz})$ & 
    14.8 & 0.2 &
    11.3 & 0.1
    \\
    $R_\Delta$ & 
    0.619 & 0.022 &
    0.913 & 0.007
    \\
    $\tau_+ \ (\times 10^{-11} \ \mathrm{s})$ &
    \multicolumn{2}{c}{$2.28^{+0.10}_{-0.09}$} &
    2.89 & 0.05
    \\
    $L_\mathrm{min}$ & 
    0.193 & 0.003 &
    0.318 & 0.001
    \\
    EM ($\mathrm{cm^{-6}\ pc}$) & 
    28.2 & 2.4 &
    4097 & 29
    \\
    dof &
    \multicolumn{2}{c}{27} & \multicolumn{2}{c}{32}
    \\
    $\chi^2/\mathrm{dof}$ &
    \multicolumn{2}{c}{3.59} & \multicolumn{2}{c}{7.65}
    \\ \hline
  \end{tabular}} \label{tab:fit_param_pol}
  \begin{tabnote}
  The errors are $1\sigma$. 
  \end{tabnote}
\end{table}
The model predictions of the polarized SED with these best-fit parameters are overlaid on the observed SED in figure \ref{fig:Pol_SED}.

It shows that our model is able to reproduce the overall features of both the intensity and polarization SEDs simultaneously.
In the best-fit model for the Perseus MC, there is a valley in the frequency dependence of the polarization fraction, and the polarization fraction reaches its minimum value at 20 GHz. 
The polarization fraction increases abruptly toward lower frequencies and approaches the asymptotic value.
The asymptotic polarization fraction is factor 5 larger than the polarization fraction in submillimeter wavebands.
The model prediction is marginally consistent with the QUIJOTE 2$\sigma$ upper limits but is slightly higher than the QUIJOTE upper limits in several frequency bands.
In the best-fit model for W43, there is a dip in the polarization fraction defined by the ratio of equation (\ref{eq:Cpol_CDE}) to equation (\ref{eq:Cabs_CDE}) from 10 to 50 GHz.
In this case, $\langle C_\nu^\mathrm{pol}\rangle$ changes sign from 13 to 30 GHz. 
Therefore, a 90 degree flip in the polarization direction in this frequency range is predicted. 
The polarization fraction below 10 GHz is factor ten larger than the polarization fraction in submillimeter wavebands.
The model prediction is marginally consistent with the QUIJOTE 2$\sigma$ upper limits but is slightly higher than the QUIJOTE upper limits in several frequency bands.

\section{Properties of the amorphous silicate dust}
\label{sec:property}

\begin{figure}[t]
  \centering
  \includegraphics{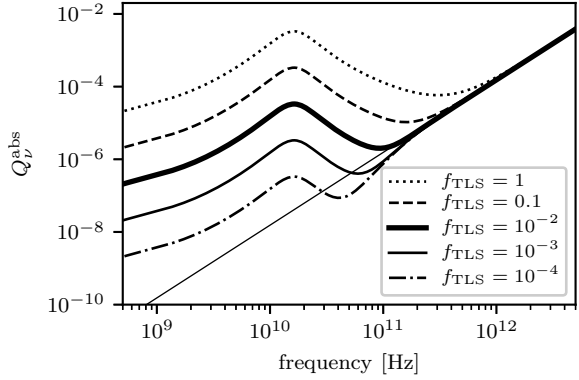}
  \caption{
  Frequency dependences of absorption efficiencies for spherical amorphous silicate dust. 
  The given values of $f_\mathrm{TLS}$ are shown in legends.
  For the other parameters, the best-fit parameter values for the Perseus MC listed in table \ref{tab:fit_param} are adopted. 
  The thin solid line is the absorption efficiency provided by the \citet{Draine_1984} model. 
  }
  \label{fig:Qabs}
\end{figure}

To reproduce the relative intensity of AME to the far infrared peak intensity, our model requires very different physical characteristics for amorphous silicate dust in comparison with amorphous silicate materials found in the laboratory.
In the laboratory, the fraction of atoms trapped in the TLS is reckoned to be of the order of $10^{-4}$.
This comes from reproducing the experimental fact that the diagnostics dominated by the TLS in the heat capacity only appears below 1 K \citep{Phillips_1987}.
On the other hand, for amorphous silicate dust, the required fraction of atoms trapped in the TLS is a few percent in order to reproduce the observed ratio of the AME peak intensity to the far infrared peak intensity with dust temperature of about 20 K.
Figure \ref{fig:Qabs} shows the frequency dependence of the absorption efficiency $Q^\mathrm{abs}_\nu$,  which is the absorption cross section normalized by the geometrical cross section, of spherical amorphous silicate dust for various TLS fractions. 
It shows that the peak value of the absorption efficiency of the resonance process of the TLS with $f_\mathrm{TLS} = 1$ is factor 5 larger than the absorption efficiency at 2 THz where the far-infrared peak appears. 
As shown in equation (\ref{eq:Inu_dust}), the thermal emission spectrum is the product of the absorption cross section and the Planck function $B_\nu(T)$. 
The ratio of the value of $B_{20{\mathrm{GHz}}}(20\ \mathrm{K})$ at the peak frequency of AME to $B_{2{\mathrm{THz}}}(20\ \mathrm{K})$ at the frequency of the far-infrared peak is about 0.0025. 
Therefore, $f_\mathrm{TLS} \sim 10^{-2}$ is required to reproduce the observed ratio of the peak intensity of AME to the far-infrared peak intensity of $10^{-4}$.
Figure \ref{fig:Qabs} also shows that the peak value of the absorption cross section of the resonance process of the TLS is two orders of magnitude less than the geometrical cross section, even in the case where $f_\mathrm{TLS} = 1$. 
It shows that this is two orders magnitude less than the absorption cross section adopted by \citet{Jones_2009}.
Therefore, his predicted SED due to the resonance process of the TLS was two orders of magnitude overestimated. 

\begin{figure}[t]
  \centering
  \includegraphics{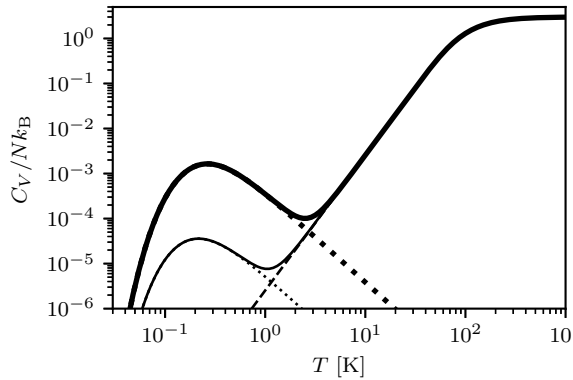}
  \caption{
  Temperature dependence of the heat capacities of the amorphous silicate dust in each MC predicted by our model.
  The contributions from the TLS are shown by a thick dotted curve for the Perseus MC and a thin dotted curve for W43. 
  The contribution from the Debye model shown by the dashed curve is the same for both MCs. 
  The thick and thin solid curves are the total heat capacities for the Perseus MC and W43, respectively.
  }
  \label{fig:CvTLS}
\end{figure}

The allowed ranges of $\Delta_0$ are narrowly limited to reproduce the bump structure in the SED.
Because of these results, the temperature dependence of the heat capacity of amorphous silicate dust has peculiar characteristics, as shown in figure \ref{fig:CvTLS}.
The heat capacity of the TLS with energy difference, $E$, is described by the Schottky heat capacity as follows:
\begin{eqnarray}
  C_V
  = \frac{E^2}{4 k_\mathrm{B} T^2}
  \mathrm{sech}^2 \left( \frac{E}{2 k_\mathrm{B} T} \right)
  .
  \label{eq:Cv_shottky}
\end{eqnarray}
Since the energy difference, $E$, has a distribution in the amorphous silicate dust, the heat capacity of the amorphous silicate dust, $C_V^\mathrm{TLS}$, is obtained by integrating over it.
The contribution of the TLS to the heat capacity of the amorphous silicate dust is then calculated as
\begin{eqnarray}
  C_V^\mathrm{TLS}
  &=&
  P_0
  \int^{\Delta_0^\mathrm{max}}_{\Delta_0^\mathrm{min}}
  \frac{d \Delta_0}{\Delta_0}
  \int^{\sqrt{(\Delta^\mathrm{max}_0)^2 - \Delta_0^2}}_0 d \Delta
  C_V
  \nonumber \\
  &=&
  P_0 \int^{\Delta_0^\mathrm{max}}_{\Delta_0^\mathrm{min}} dE
  \int^{\tau_\mathrm{max}}_{\tau_\mathrm{min}} d\tau
  \frac{C_V}{2 \tau \sqrt{1 - \tau_\mathrm{min} / \tau}}
  \nonumber \\
  &=&
  \frac{P_0 \Delta_0^\mathrm{min}}{2} \int^{1/R_\Delta}_{1} dx \ C_V
  \mathrm{asinh} \left(\sqrt{x^2 - 1} \right)
  ,
  \label{eq:Cv_TLS}
\end{eqnarray}
where $\tau$ is the relaxation time caused by the tunneling effect in equation (\ref{eq:tau_tun}) and $x \equiv E / \Delta_0^\mathrm{max}$.
There are two distinctive diagnostics compared with amorphous silicate materials in the laboratory.
The heat capacity has a bump at an extremely low temperature and is not proportional to the temperature.
This is because the allowed ranges of $\Delta_0$ are narrowly restricted.
Our model predicts that the amorphous silicate dust is composed of amorphous silicate materials, which have very distinctive characteristics compared with amorphous silicate materials found in the laboratory.

Speck et al. (\yearcite{Speck+2011}) proposed the possible forms of amorphous silicate dust in space.
If a few percent of atoms are trapped in the double-well potential caused by deformation of the crystal structure, our results are applicable to any forms of amorphous silicate dust.
The classic 10 $\micron$ amorphous silicate feature observed in the interstellar medium (\cite{Knacke+1969}; Hackwell et al. \yearcite{Hackwell+1970}) is not affected by this.

\section{Limitation of the present model and possible improvements}
\label{sec:discussion}

Although our amorphous models reproduce the observed intensity SEDs for the Perseus MC and W43, the fits were not satisfactory.
Our models underestimate the observed intensities in the frequency range from 100 GHz through 500 GHz.
The model prediction of the polarization fraction of AME is slightly higher than the QUIJOTE upper limits in several frequency bands.
The model prediction of the polarization fraction below 10 GHz is too high compared with that for submillimeter frequencies.
Possible improvements to our model for each unsatisfactory point will now be discussed.

The TLS model describes the very low temperature limit of the soft-potential model.
By fully taking into account the soft-potential model, the model SED above 100 GHz could be improved.
The TLS describes the physical behavior of amorphous materials below 1 K.
There are still deviations in the heat capacity from the Debye model in amorphous materials above 1 K, and the deviation at temperatures above 1 K is different from that below 1 K.
The plateau in the heat conductivity at $T \sim$ 1--100 K is also found in amorphous materials \citep{Zeller+1971}.
These anomalous properties of amorphous materials cannot be explained by the standard TLS model alone.
\citet{Karpov+1982} proposed the soft-potential model as a model that surpasses the standard TLS model.
The soft-potential model treats the double-well potential as the quartic function of the position of an atom.
The standard TLS model is incorporated in the soft-potential model as its very low temperature limit.
Since the typical temperature of interstellar dust is about 20 K, the physical processes beyond the standard TLS model may have a significant effect on the SED of the thermal emission from amorphous dust above 100 GHz.
\begin{figure}[t]
  \centering
  \includegraphics{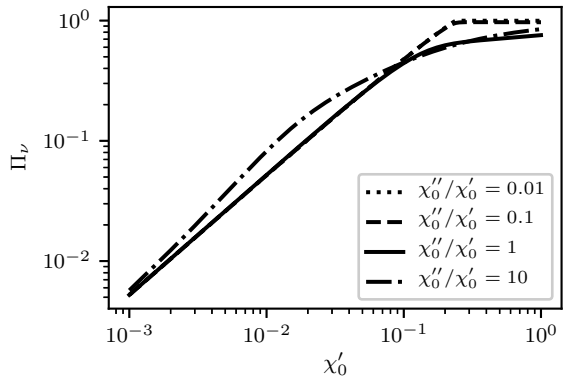}
  \caption{
  The polarization fraction as a function of the real part of the electric susceptibility ($\chi_0'$) for various ratios of the imaginary part to the real part ($\chi_0''/\chi_0'$).
  The dotted, dashed, solid, and dashed--dotted curves correspond to $\chi_0''/\chi_0'$ = 0.01, 0.1, 1, and 10, respectively. 
  The lower cut-off of $L_x$, $L_\mathrm{min}$, is set at 0.
  }
  \label{fig:Pi_chi0}
\end{figure}

\begin{figure}[t]
  \centering
  \includegraphics{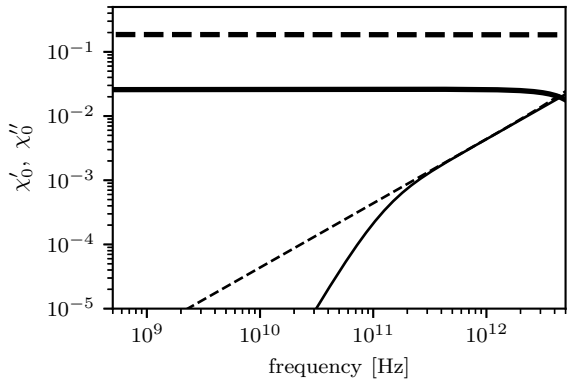}
  \caption{
  Comparison of frequency dependence of the electric susceptibility of our DCD model (solid curves) and \citet{Draine_1984} model (dashed curves). 
  Thick curves are real parts and thin curves are imaginary parts.
  }
  \label{fig:chi0_lat}
\end{figure}

One of the possibilities for reducing the model prediction of the polarization fraction in AME frequency bands is to replace the current DCD model by some other model that provides a higher value of the real part of the electric susceptibility $\chi_0$ than that of the current DCD model. 
Figure \ref{fig:Pi_chi0} shows how the polarization fraction depends on the real part of the electric susceptibility ($\chi_0'$). The plots calculated for various ratios of the imaginary part to the real part of the electric susceptibility ($\chi_0''/\chi_0'$) are shown.
The polarization fraction increases monotonically with increasing $\chi_0'$.  
As the ratio of the imaginary part to the real part decreases, the polarization fraction converges to the asymptotic value for each value of $\chi_0'$. 
This is because the polarization fraction depends only on the real part of the electric susceptibility when the imaginary part is much smaller than the real part, as shown in subsection \ref{sec:ellipsoid}.  
When the ratio is less than 0.1, the polarization fraction is proportional to $\chi_0'$ below $\chi_0'<0.2$. 
Therefore, by replacing the current  DCD model by some other model that provides a higher value of $\chi_0'$, the ellipticity required to reproduce the observed polarization fraction in submillimeter wavebands is expected to be smaller than the current model; in other words, $L_\mathrm{min}$ takes a value  closer to 1/3.
As the result, a reduction is expected in our model predictions of the polarization fraction due to the resonance process of the TLS. 
Figure \ref{fig:chi0_lat} compares the frequency dependence of the real and imaginary parts of the electric susceptibility predicted by our DCD model with that of \citet{Draine_1984} model. 
The imaginary parts of both models are identical in submillimeter wavebands.
Therefore, the intensity SEDs in submillimeter wavebands would not be changed by replacing our DCD model by the \citet{Draine_1984} model.  
The real part is an order of magnitude larger than the imaginary part in submillimeter wavebands in both models.
This shows that  the real part of the \citet{Draine_1984} model is an order of magnitude larger than that of our DCD model. 
Therefore, replacing the current DCD model by the \citet{Draine_1984} model is one possible solution to reducing the model prediction of the polarization fraction with a small change in the model prediction of the intensity SED. 
A detailed quantitative study of this possibility is beyond the scope of the present paper and will be carried out in a forthcoming paper.

We have to mention that in the range of frequencies where AME is detected the interpretation of the nature of the polarization signals is quite complex. 
The total polarized emission could be increased or decreased because of a polarized synchrotron residual component. 
In addition to this, the band pass and the beam of the telescope could mitigate the total level of polarization of AME, particularly if this happens in the frequency range where the polarization of AME is expected to change sign.

Although we have neglected the contribution of the carbonaceous dust, it is known that amorphous carbon dust is closer to the realistic form of carbonaceous dust in the interstellar medium \citep{Zubko+1996} and almost half of the mass of interstellar dust is shared by carbonaceous dust \citep{Hirashita+2009}.
Since the physical processes of the TLS are universal among the amorphous materials and independent from elemental compositions, intensity and polarization SEDs of thermal emission from amorphous dust derived in this paper are applicable to the amorphous carbon dust.
However, the physical parameters which described the TLS of the amorphous carbon dust could be different from those of the amorphous silicate dust. 
In addition, free electrons might contribute to the electric susceptibility of the amorphous carbon dust.
Further, amorphous carbon dust might have isotropic structure \citep{Draine_1984}. 
It results in the anisotropic dielectric function tensor.
Since the main scope of this paper is providing the framework to evaluate the intensity and polarization SEDs of thermal emission from amorphous dust by self-consistently taking into account the TLS model and demonstrating that this model is promising, the studies of the effect of the amorphous carbon dust on SEDs are beyond the scope of this paper and are shown in the forthcoming paper.

Although we assumed that all dust grains stayed at the same temperature, significant time variation of the temperature of the small dust grains are expected according to the stochasticity of the heating process (e.g. \cite{Draine+2001}).
In a significant fraction of time, small dust grains stay much lower temperature than that of the large dust grain which is defined by thermal equilibrium.
As shown in figure \ref{fig:SED_temp}, the relative intensity of the emission from the resonance process to the contribution from the lattice vibration becomes higher as the dust temperature becomes lower.
Therefore, quantitative studies of the stochastic heating and size distribution of the dust grains are important.

As shown in figure \ref{fig:fit_SED}, there are significant differences in the shapes of the spectra of the thermal emission from amorphous dust between the Perseus and W43 MCs. 
The variation of the spectra shape originates from the fact that the physical parameters describing the amorphous dust, such as $\Delta_0^\mathrm{max}$, $R_\Delta$, $\tau_+$, and $f_\mathrm{TLS}$, take different values for each MC (see table \ref{tab:fit_param}). 
Possible origins of the variations are now summarized.
\begin{enumerate}
  \item
  The elemental composition of amorphous silicate dust is different for the  Perseus MC and W43.
  The shape of SEDs is sensitive to the values of $\Delta_0^\mathrm{max}$ and $R_\Delta$.
  The peak frequency and width of the AME spectra due to the resonance process in the standard TLS model are defined by $\Delta_0^\mathrm{max}$ and $R_\Delta$, respectively.
  As presented in equation (\ref{eq:Delta0}), $\Delta_0$ is directly related to the tunneling parameter $\lambda$.
  The value of the tunneling parameter depends on the masses of the atoms constituting amorphous silicate dust, the width and height of the potential barrier of the double-well potential in which an atom is trapped.
  It is natural that the typical values of these parameters change if the elementary composition of amorphous silicate dust is different.
  In this paper, we have assumed the number ratio of Fe to Mg in the amorphous silicate dust to be 1 to 1 (see table \ref{tab:fixed_param}), however, that value might be different in each MC.
  Although we have not considered contributions from amorphous carbon dust in SEDs, there are several arguments that indicate their existence (e.g., \cite{Compiegne+2011}; \cite{Jones+2017}).
  It may also affect the spectral shape of AME.
  Since the elemental composition of the amorphous carbon dust is expected to be insensitive to the environmental metal abundance, the variations of the $\Delta_0^\mathrm{max}$ and $R_\Delta$ of the amorphous carbon dust among the MCs could not be  attributed to the variation of the metal abundance of the environment.
  \item
  Differences of cooling processes which solidify gas and form an amorphous dust in each MC may result in a variation of the bonding structure of the atoms in a dust and in a variation in amorphous nature of a dust particle.
  Amorphous materials are generated by rapid cooling from the liquid phase to the solid phase in laboratory.
  In interstellar space, the solidification may happen from the gas phase without passing through the liquid phase in an extremely low pressure environment.
  This could be one of the sources for which $f_\mathrm{TLS}$ takes an extremely high value compared with terrestrial  amorphous materials.
  \item
  Since the shape of the AME spectrum depends sensitively on the shape of the spectrum of free--free emission in the microwave region, which depends sensitively on the temperature of the ionized gas, it is possible that component separation between the free--free emission and AME is not sufficient.
  If the magnitude or shape of the free--free emission SED changes, the best-fit values of these parameters also change easily.
\end{enumerate}
The fact of the lack of AME in cold dense cores \citep{Tibbs+2016} could be explained by a kind of variation in the amorphous nature of the dust due to a difference in environment conditions.

\section{Conclusions}
\label{sec:conclusion}

Complete studies of the radiative processes of thermal emission from amorphous dust from the millimeter through far infrared wavebands were presented by, for the first time, self-consistently taking into account the standard TLS model.
How the intensity and the polarization SEDs respond in physical parameters characterizing the standard TLS model was shown.
The amorphous model could reproduce very well the observed SEDs from AME up to the far-infrared feature.
In our models, AME is originated mainly from the resonance emission of the TLS of large grains.
The amorphous model is able to explain AME without introducing new species.
Simultaneous fitting of the polarization and intensity SED for the Perseus MC and W43 were also performed.
Since there is no definite detection of polarization emission from AME, the adopted polarization intensities in the AME frequency range were upper limits.
The polarization intensities measured by Planck at 143, 217, and 353 GHz were also included.
The amorphous model could reproduce the overall observed feature of the intensity and polarization SEDs of the Perseus MC and W43.
However, the model prediction of the polarization fraction of AME was slightly higher than the QUIJOTE upper limits in several frequency bands.
Possible improvements to our model to resolve this problem were proposed in the previous section.
Our model predicts that amorphous silicate dust have very different physical characteristics compared with amorphous silicate materials found in the laboratory.
We have shown that thermal emission from amorphous dust is an attractive alternative possibility as the origin of AME.

\bigskip
\begin{ack}
We thank Tetsuo Yamamoto for helpful discussions throughout the course of this work. 
We thank the referee, Itsuki Sakon, for constructive comments.
We thank T.J. Mahoney for revising the English of the draft.
MN acknowledges support from the Graduate Program on Physics for the Universe (GP-PU), Tohoku University. 
FP acknowledges the European Commission and the MINECO.
This work is partially supported by MEXT KAKENHI Grant Number 18H05539 and MEXT KAKENHI Grant Number 18H01250. 
This project has been partially funded by the European Union's Horizon 2020 research and innovation programme under grant agreement number 687312 (RADIOFOREGROUNDS), and by the SPACE IR MISSIONS II project under grants agreements ESP2015-65597-C4-4-R and ESP2017-86852-C4-2-R, respectively.
MH would like to express his sincere condolences to Prof. Tsai An-Pang who was the world authority on quasicrystal and passed away in May 2019 at age of 60. In the course of this study, as the resident who lived in the same apartment by chance, MH received fruitful comments on the study and great support in the private life.
\end{ack}

\appendix
\section{Standard TLS model}
\label{sec:TLSmodel}

The basics of the standard TLS model are as follows.
The ground states confined in each harmonic potential are denoted by $\varphi_1$ and $\varphi_2$, respectively.
They are defined by the Schr\"{o}dinger equations as
\begin{eqnarray}
  H_i | \varphi_i \rangle
  &=&
  \left(-\frac{\hbar^2}{2m} \frac{d^2}{dx^2}  + V_i \right) | \varphi_i \rangle
  =
  \epsilon_i | \varphi_i \rangle
  ,
  \label{eq:SchrodingerEq_phii}
\end{eqnarray}
where $i$ runs from 1 to 2, $\epsilon_i$ is the energy of the ground state, and $m$ is the mass of an atom. 
The wave functions  $\langle x | \varphi_1 \rangle$ and $\langle x | \varphi_2 \rangle$ are localized around the bottom of the potentials $V_1$ and $V_2$, respectively.
In other words, there is almost no overlap between $\langle x | \varphi_1 \rangle$ and $\langle x | \varphi_2 \rangle$; thus $\langle \varphi_1 | \varphi_2 \rangle$ vanishes in practice; i.e., 
\begin{eqnarray}
  \langle \varphi_1 | \varphi_2 \rangle
  = \int dx' \langle \varphi_1 | x' \rangle
             \langle x' | \varphi_2 \rangle
  \simeq 0
  .
\end{eqnarray}
The Hamiltonian of the standard TLS model is described by
\begin{eqnarray}
  H = H_1 - V_1 + V = H_2 - V_2 + V
  .
  \label{eq:Hamiltonian_whole}
\end{eqnarray}
The ground state, $\psi_1$, and the first excited state, $\psi_2$, confined in the double-well potential satisfy the following Schr\"{o}dinger equations:
\begin{eqnarray}
  E_k | \psi_k \rangle
  = H  |\psi_k \rangle
  ,
  \label{eq:eigenEq_psi_i}
\end{eqnarray}
where $E_1$ and $E_2$ are the energy eigenvalues of the ground state and the first excited state, respectively.
Under the standard TLS approximation, these states are represented in the form $|\psi_i \rangle = \sum_{j=1,2} c_{ij} | \varphi_j \rangle$.
Using equation (\ref{eq:Hamiltonian_whole}), equation (\ref{eq:eigenEq_psi_i}) is expressed in matrix form:
\begin{eqnarray}
  &&
  E_i 
  \left( \!\!\!\!\!
  \begin{array}{c} c_{i1} \\ c_{i2} \end{array} \!\!\!\!\! \right)
  =
  H
  \left(\!\!\!\!\!
  \begin{array}{c} c_{i1} \\ c_{i2} \end{array} \!\!\!\!\!\right)
  ,
  \label{eq:eigenEq_psi_i_matrx}
  \\
  &&
  H
  =
  \left(\!\!\!\!\! \begin{array}{cc}
  \epsilon_1 \mathalpha{+} \langle \varphi_1 |
  (V\mathalpha{-}V_1) | \varphi_1 \rangle \!\!\!\! &
  \langle \varphi_1 | H | \varphi_2 \rangle \\
  \langle \varphi_2 | H | \varphi_1 \rangle & \!\!\!\!
  \epsilon_2 \mathalpha{+} \langle \varphi_2 |
  (V\mathalpha{-}V_2) | \varphi_2 \rangle
  \end{array} \!\!\!\!\!\right)
  .
\end{eqnarray}
Since each $\varphi_i$ is located at the bottom of $V_i$, $\langle \varphi_i |(V-V_i)| \varphi_i \rangle \ll \epsilon_i$, and the diagonal elements are approximated by $\epsilon_i$.
Two variables, $\Delta \equiv \epsilon_2 - \epsilon_1$ and 
$\Delta_0 \equiv 2 \langle \varphi_1 | H | \varphi_2 \rangle = 2 \langle \varphi_2 | H | \varphi_1 \rangle$, are introduced to characterize the TLS.
$\Delta$ is the energy difference between the two states located at each minimum of the double-well potential and characterizes the degree of asymmetry of the potential.
$\Delta_0$ is the parameter that characterizes the degree of the cross correlation between the states located in two minima and can be approximated by
\begin{eqnarray}
  \Delta_0
  = 
  \hbar \Omega e^{- \lambda}
  ,
  \label{eq:Delta0}
\end{eqnarray}
where $\hbar \Omega$ is the order of $\epsilon_1$ and $\epsilon_2$.
Note that $\lambda$ is used for the tunneling parameter in this section. 
By shifting the meaningless zero level of the energy eigenvalues, $H$ is rewritten with these two parameters as
\begin{eqnarray}
  H
  =
  \frac{1}{2}
  \left( \begin{array}{cc}
  -\Delta & \Delta_0 \\
  \Delta_0 & \Delta
  \end{array} \right)
  .
  \label{eq:H_matrix}
\end{eqnarray}
The energy eigenvalues $E_i$ are obtained by deducing the eigenvalues of the matrix $H$, written in equation (\ref{eq:H_matrix}) as,
\begin{eqnarray}
  E_1 &=& -\frac{E}{2}
  ,
  \label{eq:E1} \\
  E_2 &=& \frac{E}{2}
  ,
  \label{eq:E2} \\
  E &\equiv& \sqrt{\Delta^2 + \Delta_0^2}
  ,
  \label{eq:E}
\end{eqnarray}
where $E$ is the energy splitting of the TLS.
By normalizing the states $\psi_1$ and $\psi_2$ as
$\langle \psi_1 | \psi_1 \rangle = \langle \psi_2 | \psi_2 \rangle = 1$, the expansion coefficients are expressed by using a single parameter as 
$c_{11} = \cos \theta$, $c_{12} = -\sin \theta$,
$c_{21} = \sin \theta$, and $c_{22} = \cos \theta$
where $\cos 2 \theta \equiv \Delta / E$, $\sin 2 \theta \equiv \Delta_0 / E$.
Therefore, $| \psi_1 \rangle$ and $| \psi_2 \rangle$ are represented by
\begin{eqnarray}
  | \psi_1 \rangle
  &=&
  \cos \theta | \varphi_1 \rangle - \sin \theta | \varphi_2 \rangle
  ,
  \label{eq:psi1} \\
  | \psi_2 \rangle
  &=&
  \sin \theta | \varphi_1 \rangle + \cos \theta | \varphi_2 \rangle
  .
  \label{eq:psi2}
\end{eqnarray}
Suppose the two energy eigenstates set up a complete system. 
Then an arbitrary state of the TLS $|\psi \rangle$ can be described by
\begin{eqnarray}
  | \psi \rangle
  = a_1(t) | \psi_1 \rangle + a_2(t) | \psi_2 \rangle
  ,
  \label{eq:psi}
\end{eqnarray}
where $a_1$ and $a_2$ are time-dependent complex numbers and satisfy the normalization condition ($|a_1|^2 + |a_2|^2 = 1$).
Furthermore, the identity operator can be defined as 
$\hat{\sigma}_I \equiv | \psi_1 \rangle \langle \psi_1 | + | \psi_2 \rangle \langle \psi_2 |$.

The following operators are useful for seeing the physical behavior of the TLS, such as
\begin{eqnarray}
  \hat{\sigma}_+
  &\equiv&
  | \psi_2 \rangle \langle \psi_1 |
  ,
  \label{eq:sigma_+}
  \\
  \hat{\sigma}_-
  &\equiv&
  | \psi_1 \rangle \langle \psi_2 |
  ,
  \label{eq:sigma_-}
  \\
  \hat{\sigma}_w
  &\equiv&
  | \psi_1 \rangle \langle \psi_1 | - | \psi_2 \rangle \langle \psi_2 |
  .
  \label{eq:sigma_w}
\end{eqnarray}
In order to clarify the physical meanings of these operators, let them act on $|\psi \rangle$.
We then obtain
\begin{eqnarray}
  \hat{\sigma}_+ | \psi \rangle &=& a_1 | \psi_2 \rangle
  ,
  \\
  \hat{\sigma}_- | \psi \rangle &=& a_2 | \psi_1 \rangle ,
  \\
  \hat{\sigma}_w | \psi \rangle &=& a_1 | \psi_1 \rangle - a_2 | \psi_2 \rangle
  .
\end{eqnarray}
The expectations of each operator, $u_\pm$ and $w$, can be calculated as,
\begin{eqnarray}
  u_+ &=& \langle \psi | \hat{\sigma}_+ | \psi \rangle = a_1 a_2^*
  ,
  \\
  u_- &=& \langle \psi | \hat{\sigma}_- | \psi \rangle = a_1^* a_2
  ,
  \\
  w &=& \langle \psi | \hat{\sigma}_w | \psi \rangle = a_1 a_1^* - a_2 a_2^*
  .
\end{eqnarray}
Therefore, $\hat{\sigma}_\pm$ are something like ladder operators and $\hat{\sigma}_w$ measures a difference of the probabilities of finding an atom in each state.
The operator $\hat{\sigma}_+$ represents the excitation of the ground state $\psi_1$ to the excited state $\psi_2$.
The operator $\hat{\sigma}_-$ represents the downward transition from the excited state to the ground state.

The interaction Hamiltonian between the TLS and an electromagnetic field, $H'$, can be written,
\begin{eqnarray}
  H' = -q \bm{r} \cdot \bm{\mathcal{E}}_\mathrm{local} 
  ,
  \label{eq:Hamiltonian_dipole}
\end{eqnarray}
where $q$ is the charge of an atom trapped in the double-well potential, $\bm{r}$ is its position vector, and $\bm{\mathcal{E}}_\mathrm{local}$ is the local electric field at the position of the atom.
The magnetic effect is negligible since the velocity of the atom is much smaller than the speed of light.
Evolution of the atomic state $\psi$ caused by the perturbation is described by the following Schr\"{o}dinger equation,
\begin{eqnarray}
  i \hbar \frac{\partial |\psi \rangle}{\partial t}
  = \left(H + H' \right) |\psi \rangle 
  .
  \label{eq:SchrodingerEq_psi}
\end{eqnarray}
The electric dipole moment arising from the TLS, $\bm{d}_\mathrm{TLS}$, is given by,
\begin{eqnarray}
  \bm{d}_\mathrm{TLS}
  &=&
  \langle \psi |q \bm{r}| \psi \rangle
  \nonumber \\
  &\simeq&
  \left[
    -\left( a_1 a_1^* - a_2 a_2^* \right) \cos 2 \theta
    -\left( a_1 a_2^* + a_1^* a_2 \right) \sin 2 \theta
  \right] \bm{d}_0
  ,
  \label{eq:dipole} \\
  \bm{d}_0
  &\equiv&
  -\langle \varphi_1 |q \bm{r}| \varphi_1 \rangle
  =\langle \varphi_2 |q \bm{r}| \varphi_2 \rangle
  .
  \label{eq:dipole0}
\end{eqnarray}
The cross terms,
$\langle \varphi_1 |q \bm{r}| \varphi_2 \rangle$ and
$\langle \varphi_2 |q \bm{r}| \varphi_1 \rangle$,
are neglected in comparison with the diagonal terms.
We can choose the origin of the coordinate to realize equation (\ref{eq:dipole0}) without losing generality.
$\bm{d}_0$ is the electric dipole moment for the state located at the minimum of the potential $V_2$.
The time evolution equations for $a_1$ and $a_2$ are led by equation (\ref{eq:SchrodingerEq_psi}),
\begin{eqnarray}
  \frac{da_1}{dt}
  &=&
  i a_1 \left(+\frac{\omega_0}{2} + \Omega_0 \cos 2 \theta \right)
  -i a_2 \Omega_0 \sin 2 \theta
  ,
  \label{eq:da1/dt} \\
  \frac{da_2}{dt}
  &=&
  i a_2 \left(-\frac{\omega_0}{2} + \Omega_0 \cos 2 \theta \right)
  -i a_1 \Omega_0 \sin 2 \theta
  ,
  \label{eq:da2/dt}
\end{eqnarray}
where $\hbar \omega_0 \equiv E$ and $\hbar \Omega_0 \equiv \bm{d}_0 \cdot \bm{\mathcal{E}}_\mathrm{local}$.
Defining a Bloch vector $\bm{R} \equiv (u,\ v,\ w) \equiv (a_1a_2^* + a_1^*a_2,\ -i(a_1a_2^*-a_1^*a_2),\ a_1 a_1^* - a_2 a_2^*)$, each component of $\bm{R}$ satisfies the following equations:
\begin{eqnarray}
  \frac{du}{dt}
  &=&
  -v (\omega_0 - 2 \Omega_0 \cos 2 \theta)
  ,
  \label{eq:du/dt} \\
  \frac{dv}{dt}
  &=&
  u (\omega_0 - 2 \Omega_0 \cos 2 \theta) + 2 w \Omega_0 \sin 2 \theta
  ,
  \label{eq:dv/dt} \\
  \frac{dw}{dt}
  &=&
  -2 v \Omega_0 \sin 2 \theta
  .
  \label{eq:dw/dt}
\end{eqnarray}
Equations (\ref{eq:du/dt})--(\ref{eq:dw/dt}) are called Bloch equations
\footnote
{
By introducing a vector
$\bm{\Omega} \equiv
(-2 \Omega_0 \sin 2 \theta,\ 0,\ \omega_0 -2 \Omega_0 \cos 2 \theta)$,
equations (\ref{eq:du/dt})--(\ref{eq:dw/dt}) are rewritten as
$d\bm{R} / dt = \bm{\Omega} \times \bm{R}$.
}.
Equations for $u_{\pm}$ are led by equations (\ref{eq:du/dt}) and (\ref{eq:dv/dt}) as,
\begin{eqnarray}
  \frac{du_\pm}{dt}
  =
  \pm i u_\pm (\omega_0 - 2 \Omega_0 \cos 2 \theta)
  \pm i w \Omega_0 \sin 2 \theta
  .
  \label{eq:du+-/dt}
\end{eqnarray}
Due to the spontaneous transition, $w$ is relaxed to the instantaneous thermal equilibrium state $\bar{w}(t)$.
The energy levels of the TLS are shifted as
$E_1 \rightarrow E_1 +({\bm d}_0 \cos 2 \theta) \cdot \bm{\mathcal{E}}_\mathrm{local}$ and
$E_2 \rightarrow E_2 -({\bm d}_0 \cos 2 \theta) \cdot \bm{\mathcal{E}}_\mathrm{local}$
owing to the interaction of the atom with the local electric field.
As a result, the population of the thermal equilibrium states changes to $\bar{w}(t)$.
This is the instantaneous thermal equilibrium state of the population.
By introducing the relaxation time $\tau_w$, equation (\ref{eq:dw/dt}) is modified as,
\begin{eqnarray}
  \frac{dw}{dt}
  &=&
  -2 v \Omega_0 \sin 2 \theta - \frac{w - \bar{w}(t)}{\tau_w}
  .
  \label{eq:BlochEq_w}
\end{eqnarray}
$u_+$ represents the excitation of the atom in the TLS by the absorption of the electromagnetic wave.
$u_-$ represents the transition from the excited state to the ground state stimulated by the electromagnetic wave.
The relaxation of the states to the original states is taken into account in the evolution equations of $u_{\pm}$ by introducing the phase relaxation time $\tau_+$ such that
\begin{eqnarray}
  \frac{du_\pm}{dt}
  &=&
  \pm i u_\pm (\omega_0 - 2 \Omega_0 \cos 2 \theta)
  \pm i w \Omega_0 \sin 2 \theta - \frac{u_\pm}{\tau_+}
  .
  \label{eq:BlochEq_u+-}
\end{eqnarray}
When the local electric field carried by the electromagnetic wave is weak enough, the Bloch equations (\ref{eq:BlochEq_u+-}) and (\ref{eq:BlochEq_w}) can be treated perturbatively.
The zeroth order solutions are given by
\begin{eqnarray}
  u_\pm^{(0)}
  &=&
  \frac{1}{2} \mathrm{sech} \left( \frac{E}{2 k_\mathrm{B} T} \right)
  e^{\pm i (\omega_0 t + \delta)}
  ,
  \\
  w^{(0)}
  &=&
  \tanh \left( \frac{E}{2 k_\mathrm{B} T} \right)
  ,
\end{eqnarray}
where $k_\mathrm{B}$ is the Boltzmann constant.
Since the zeroth order states are the thermal equilibrium states, the population of each level is given by
$|a_1^{(0)}|^2 = 1 / \{\exp [-E / (k_\mathrm{B} T)] + 1\}$ and
$|a_2^{(0)}|^2 = 1 / \{\exp [E / (k_\mathrm{B} T)] + 1\}$, respectively.
Since the phase coefficients $\exp (\pm i \delta)$ take random values through the whole amorphous material, we may assume $u^{(0)}_\pm = 0$.

In the first order of perturbations, the Bloch equations are written
\begin{eqnarray}
  \frac{du_\pm^{(1)}}{dt}
  &=&
  \pm i \omega_0 u_\pm^{(1)}
  \pm i w^{(0)} \Omega_0 \sin 2 \theta - \frac{u_\pm^{(1)}}{\tau_+}
  ,
  \\
  \frac{dw^{(1)}}{dt}
  &=&
  - \frac{w^{(1)} - \bar{w}^{(1)}}{\tau_w}
  .
\end{eqnarray}
 $\bar{w}^{(1)}$ is estimated by expanding $\bar{w}$ with $E$,
\begin{eqnarray}
  \bar{w}^{(1)}
  &=&
  \frac{\partial \bar{w}}{\partial E}
  \left[ 
  -2 (\bm{d}_0 \cos 2 \theta) \cdot \bm{\mathcal{E}}_\mathrm{local} 
  \right]
  \nonumber \\
  &=&
  -\frac{\bm{d}_0 \cdot \bm{\mathcal{E}}_\mathrm{local} \cos 2 \theta}
  {k_\mathrm{B} T}
  \mathrm{sech} \left( \frac{E}{2 k_\mathrm{B} T} \right)
  .
  \label{eq:w_bar1}
\end{eqnarray}
By decomposing the incident electric field into the Fourier spectrum, 
$\bm{\mathcal{E}}_\mathrm{local}(t) = \int d\omega \hat{\bm{\mathcal{E}}}_\mathrm{local}(\omega) e^{-i \omega t}$,
the first order solutions of the Bloch equations are:
\begin{eqnarray}
  \hat{u}_\pm^{(1)}
  &=&
  \pm i \frac{\tau_+}{\hbar}
  \frac{\bm{d}_0 \cdot \hat{\bm{\mathcal{E}}}_\mathrm{local} \sin 2 \theta}
  {1 + i (\omega \mp \omega_0) \tau_+}
  \tanh \left( \frac{E}{2 k_\mathrm{B} T} \right)
  ,
  \label{eq:u+-_hat} \\
  \hat{w}^{(1)}
  &=&
  -\frac{\bm{d}_0 \cdot \hat{\bm{\mathcal{E}}}_\mathrm{local} \cos 2 \theta}
  {k_\mathrm{B} T}
  \frac{1}{1 - i \omega \tau_w}
  \mathrm{sech} \left( \frac{E}{2 k_\mathrm{B} T} \right)
  .
  \label{eq:w_hat}
\end{eqnarray}

To obtain the absorption coefficient of the amorphous material against electromagnetic waves, the electric susceptibilities based on the standard TLS model are deduced.
Before going into detail on each physical process, we have to model the distributions of $\Delta$ and $\Delta_0$.
\citet{Anderson+1972} and \citet{Phillips_1972} proposed that the probability of finding $\lambda$ and $\Delta$ at some value is uniform since the possible range of these variables is narrowly limited.
The distribution function of $\Delta$ and $\Delta_0$, $f(\Delta_0,\ \Delta)$ is then given by:
\begin{eqnarray}
f(\Delta_0,\ \Delta) d\Delta_0 d\Delta
&=&
P_0 d\lambda d\Delta
=
P_0 \left| \frac{\partial \lambda}{\partial \Delta_0} \right|
d\Delta_0 d\Delta
,
\nonumber \\
f(\Delta_0,\ \Delta)
&=&
\frac{P_0}{\Delta_0}
,
\label{eq:PDF}
\end{eqnarray}
where $f(\Delta_0,\ \Delta) d\Delta_0 d\Delta$ provides the number density of the atoms trapped in the TLS from $\Delta_0$ to $\Delta_0 + d\Delta_0$ and  from $\Delta$ to $\Delta + d\Delta$, and $P_0$ is the constant providing the number density of atoms trapped in the TLS, $n_\mathrm{TLS}$, which is deduced by integrating the distribution function over $d\Delta_0$ and $d\Delta$ as,
\begin{eqnarray}
n_\mathrm{TLS}
&=&
P_0
\int^{\Delta^\mathrm{max}_0}_{\Delta^\mathrm{min}_0} \frac{d\Delta_0}{\Delta_0}
\int^{\sqrt{(\Delta^\mathrm{max}_0)^2 - \Delta_0^2}}_0 d\Delta
\nonumber \\
&=&
P_0\Delta_0^\mathrm{max}
\left[
\ln \left( \frac{\sqrt{1 - R_\Delta^2} + 1}{R_\Delta} \right)
-\sqrt{1 - R_\Delta^2}
\right]
,
\label{eq:nTLS} \\
R_\Delta
&\equiv&
\frac{\Delta_0^\mathrm{min}}{\Delta_0^\mathrm{max}}
.
\label{eq:r_Del0}
\end{eqnarray}
$\Delta_0^\mathrm{max}$ and $\Delta_0^\mathrm{min}$ are introduced to avoid divergence of the probability distribution function.
$\Delta_\mathrm{max}$ and $\Delta_0^\mathrm{min}$ are related to each other through the maximum energy splitting of the TLS 
as $E_\mathrm{max}^2 = \Delta_\mathrm{min}^2 + (\Delta_0^\mathrm{max})^2 = \Delta_\mathrm{max}^2 + (\Delta_0^\mathrm{min})^2$ 
(see equation (\ref{eq:E})).
We treat $\Delta_\mathrm{max}$ as a dependent variable of $\Delta_0^\mathrm{min}$.
For simplicity, we set $\Delta_\mathrm{min}$ to zero.

The expectations $u_\pm$ represent the transition between the TLS due to absorption and emission of the electromagnetic wave.
These processes refer to the resonance transition.
We derive the electric susceptibility due to the resonance transition.
The electric dipole moment caused by the resonance transition, $\bm{d}_\mathrm{res}$, stimulated by an electromagnetic wave of angular frequency $\omega$ can be written as
\begin{eqnarray}
  \bm{d}_\mathrm{res}
  &\simeq&
  -\left(\hat{u}_+^{(1)} + \hat{u}_-^{(1)}\right) \bm{d}_0 \sin 2 \theta
  \nonumber \\
  &=&
  -i \frac{\tau_+}{\hbar}
  \bm{d}_0 \cdot \hat{\bm{\mathcal{E}}}_\mathrm{local}
  \left( \frac{\Delta_0}{E} \right)^2
  \tanh \left( \frac{E}{2 k_\mathrm{B} T} \right)
  \nonumber \\ && \times \
  \left[
    \frac{1}{1 + i(\omega - \omega_0) \tau_+}
  - \frac{1}{1 + i(\omega + \omega_0) \tau_+}
  \right]
  \bm{d}_0
  .
  \label{eq:dipole_res}
\end{eqnarray}
The electric polarization $\bm{P}_\mathrm{res}$ is calculated by averaging the electric dipole moment over the solid:
\begin{eqnarray}
  \bm{P}_\mathrm{res}
  =\frac{1}{\mathcal{V}} \sum_i \bm{d}_\mathrm{res}^i
  =\int d\Delta_0 \int d\Delta f(\Delta_0,\ \Delta) \bm{d}_\mathrm{res}
  ,
  \label{eq:P_res}
\end{eqnarray}
where $\mathcal{V}$ is the volume of an amorphous material. 
In generally, the electric polarization $\bm{P}$ of an isotropic and spherical particle is related to the external electric field $\bm{\mathcal{E}}_\mathrm{ext}$,
\begin{eqnarray} 
  \bm{P} 
  = \chi_0 \bm{\mathcal{E}}_\mathrm{ext}
  , 
\end{eqnarray}
where $\chi_0$ is the electric susceptibility for the response to an external electric field.
In spherical dielectric material, a local electric field equals an externally applied field (see appendix \ref{sec:localF}).
Thus, $\chi_0^\mathrm{res}$ is given by,
\begin{eqnarray}
  \chi_0^\mathrm{res}
  &=&
  \frac{\bm{P}_\mathrm{res} \cdot \hat{\bm{\mathcal{E}}}_\mathrm{ext}}
  {| \hat{\bm{\mathcal{E}}}_\mathrm{ext} |^2}
  = 
  \frac{\bm{P}_\mathrm{res} \cdot \hat{\bm{\mathcal{E}}}_\mathrm{local}}
  {| \hat{\bm{\mathcal{E}}}_\mathrm{local} |^2}
  .
\end{eqnarray}
By assuming that the directions of $\bm{d}_0$ relative to the local electric field $\hat{\bm{\mathcal{E}}}_\mathrm{local}$ are randomly distributed, the average of $(\bm{d}_0 \cdot \hat{\bm{\mathcal{E}}}_\mathrm{local})^2$ becomes 
$|\bm{d}_0|^2 |\hat{\bm{\mathcal{E}}}_\mathrm{local}|^2 / 3$.
Then, the electric susceptibility described by equation (\ref{eq:chi0_res}) is obtained.

The expectation $w$ relaxes to the instantaneous thermal equilibrium value.
We describe how the relaxation process contributes to the electric susceptibility.
The electric dipole moment due to the relaxation process is written as,
\begin{eqnarray}
  \bm{d}_\mathrm{rel}
  &\simeq&
  -\hat{w}^{(1)} \bm{d}_0 \cos 2 \theta
  \nonumber \\
  &=&
  \frac{\bm{d}_0 \cdot \hat{\bm{\mathcal{E}}}_\mathrm{local}}{k_\mathrm{B} T}
  \left( \frac{\Delta}{E} \right)^2
  \frac{1}{1 - i \omega \tau_w}
  \mathrm{sech} \left( \frac{E}{2 k_\mathrm{B} T} \right) \bm{d}_0
  .
  \label{eq:dipole_rel}
\end{eqnarray}
There are two main relaxation processes.
One is quantum tunneling in which an atom passes through the potential barrier by the quantum effect.
The other is hopping where an atom climbs over the barrier by gaining enough energy due to thermal fluctuation.

The tunneling relaxation time $\tau_\mathrm{tun}$ was deduced by \citet{Phillips_1972} as
\begin{eqnarray}
  \tau_\mathrm{tun}^{-1}
  &=&
  \left( \frac{\gamma_\mathrm{l}^2}{c_\mathrm{l}^5}
	+\frac{2 \gamma_\mathrm{t}^2}{c_\mathrm{t}^5}\right)
	\frac{\omega_0 \Delta_0^2}{2 \pi \rho \hbar^3}
	\mathrm{coth} \left( \frac{E}{2 k_\mathrm{B} T} \right)
  ,
  \label{eq:tau_tun}
\end{eqnarray}
where $\gamma_\mathrm{t(l)}$ and $c_\mathrm{t(l)}$ are the elastic dipole and sound velocity for the transverse (longitudinal) waves, respectively.
$\rho$ is the mass density.
Typical values of physical variables of amorphous silicate material found in laboratory experiments are listed in table \ref{tab:fixed_param} (where $c_\mathrm{l}^{-5} \ll 2c_\mathrm{t}^{-5}$).
Then, the complex susceptibility for the tunneling relaxation $\chi_0^\mathrm{tun}$ is obtained as equation (\ref{eq:chi0_tun}).

The hopping relaxation time $\tau_\mathrm{hop}$ is given by Arrhenius equation as,
\begin{eqnarray}
  \tau_\mathrm{hop}^{-1}
  &=&
  \frac{1}{\tau^0_\mathrm{hop}} \exp \left(-\frac{V_0}{k_\mathrm{B} T} \right)
  ,
  \label{eq:tau_hop}
\end{eqnarray}
where the values of $\tau^0_\mathrm{hop}$ are defined by the physical characteristics of each amorphous material.
The relaxation time scale of the hopping is sensitive to the height of the potential barrier $V_0$, which must vary in value across a single dust grain.
The probability density function $g(V_0)$ for $V_0$ is introduced.
\citet{Bosch_1978} proposed the Gaussian distribution function of $g(V_0)$ as
\begin{eqnarray}
  g(V_0)
  &=&
  \left\{ \begin{array}{cc}
  C_{V_0} \exp \left[-\left( \frac{V_0 - V_m}{V_\sigma} \right)^2 \right] ,
  & V_0 > V_\mathrm{min} ; \\[2ex]
  0 , & V_0 < V_\mathrm{min} ;
  \end{array}\right
  .
  \label{eq:f_V0} \\
  C_{V_0}
  &=&
  \frac{2}{V_\sigma \sqrt{\pi}}
  \left[ \mathrm{Erf} \left( \frac{V_m - V_\mathrm{min}}{V_\sigma} \right)
  + 1 \right]^{-1}
  ,
  \\
  \mathrm{Erf}(x)
  &\equiv&
  \int^x_0 dt \ e^{-t^2}
  .
  \label{eq:Erf}
\end{eqnarray}
By taking into account the distribution of $V_0$, the complex susceptibilities for the hopping relaxation $\chi_0^\mathrm{hop}$ is obtained as equation (\ref{eq:chi0_hop}).

\section{Extension of the Clausius-Mossotti relation for an ellipsoidal particle}
\label{sec:localF}
 
Consider a homogeneous ellipsoid located in a uniform electric field $\bm{\mathcal{E}}_\mathrm{ext}^i$,  whose direction is parallel to the $i$th semi-axis of the ellipsoidal particle. 
An electric polarization $\bm{P}_i$ arising from $\bm{\mathcal{E}}_\mathrm{ext}^i$ aligns in the same direction. 
$\bm{P}_i$ is given as (e.g., \cite{Bohren+1983}), 
\begin{eqnarray}
  \bm{P}_i 
  = \frac{1}{4\pi} \frac{\varepsilon - 1}{1 + L_i (\varepsilon - 1)} \bm{\mathcal{E}}_\mathrm{ext}^i
  \equiv \chi_0^i \bm{\mathcal{E}}_\mathrm{ext}^i
  ,
  \label{eq:Pi}
\end{eqnarray}
where $L_i$ is a geometrical factor defined by equation (\ref{eq:Li}), $\varepsilon$ is the electric permittivity of the particle, and $\chi_0^i$ is the electric susceptibility  along the $i$th semi-axis of the ellipsoid for the external field.

The local electric field $\bm{\mathcal{E}}_\mathrm{local}$, based on Lorentz's approach, is the sum of the external field and the electric field generated by the electric polarization (see \cite{Kittel}), 
\begin{eqnarray}
  \bm{\mathcal{E}}_\mathrm{local}
  = \bm{\mathcal{E}}_\mathrm{ext}
  + \bm{\mathcal{E}}_1
  + \bm{\mathcal{E}}_2
  + \bm{\mathcal{E}}_3
  ,
  \label{eq:Elocal}
\end{eqnarray}
where $\bm{\mathcal{E}}_1$ is the depolarization field generated from a surface charge density, $\bm{\mathcal{E}}_2$ is the electric field produced by a surface electric charge density on a virtual spherical cavity, and $\bm{\mathcal{E}}_3$ is the electric field created by dipole moments inside the cavity. 
$\bm{\mathcal{E}}_1$ is related to the electric polarization according to 
\begin{eqnarray}
  \bm{\mathcal{E}}_1^i = - 4 \pi L_i \bm{P}_i
  .
  \label{eq:E1i}
\end{eqnarray}
$\bm{\mathcal{E}}_2$ is expressed by the electric polarization as, 
\begin{eqnarray}
  \bm{\mathcal{E}}_2^i = \frac{4}{3} \pi \bm{P}_i
  .
  \label{eq:E2i}
\end{eqnarray}
In amorphous material, it is expected that the position of each atom is completely random, and that the electric fields from dipole moments cancel each other out; therefore, $\bm{\mathcal{E}}_3 = 0$. 
Using equations (\ref{eq:Pi})--(\ref{eq:E2i}), $\bm{\mathcal{E}}_\mathrm{local}$ may be calculated:  
\begin{eqnarray}
  \bm{\mathcal{E}}_\mathrm{local}^i
  =
  \frac{1}{3}
  \frac{\varepsilon + 2}{1 + L_i(\varepsilon - 1)}
  \bm{\mathcal{E}}_\mathrm{ext}^i 
  .
  \label{eq:Elocal_i}
\end{eqnarray}
In a spherical particle, the local field is equal to the external field because $L_i = 1/3$ (see \cite{Kittel}), whereas in an ellipsoidal particle, the local field differs from the external field. 
Since an electric field acting each atom is the local field, the electric polarization can be described as, 
\begin{eqnarray}
  \bm{P}_i
  = \left(\sum_j N_j \alpha_j \right) \bm{\mathcal{E}}_\mathrm{local}^i
  ,
  \label{eq:Pi_local}
\end{eqnarray}
where $\alpha_j$ is the polarizability of each atom $j$ and $N_j$ is the concentration. 
From equations (\ref{eq:Pi}), (\ref{eq:Elocal_i}) and (\ref{eq:Pi_local}), we may obtain the following relation: 
\begin{eqnarray}
  \sum_j N_j \alpha_j 
  = \frac{3}{4 \pi} \frac{\varepsilon - 1}{\varepsilon + 2}
  .
  \label{eq:CM_relation}
\end{eqnarray}
Equation (\ref{eq:CM_relation}) is the Clausius-Mossotti relation, which is satisfied regardless of the shape of the particle. 
In other words, the electric permittivity is a physical parameter independent of the particle shape.
This equation relates microscopic physical parameters to macroscopic physical parameters.
From equation (\ref{eq:Elocal_i}), we can see that the local field equals the external field for a spherical particle.
Therefore, $\sum_j N_j \alpha_j$ is equal to $\chi_0$ which is the electric susceptibility of a spherical particle for the external electric field. 
From equation (\ref{eq:CM_relation}) we get
\begin{eqnarray}
  \chi_0
  = \frac{3}{4 \pi} \frac{\varepsilon - 1}{\varepsilon + 2}
  .
  \label{eq:chi0_sphere}
\end{eqnarray}
We can derive the relation between $\chi_0^i$ and $\chi_0$ from equations (\ref{eq:Pi}) and (\ref{eq:chi0_sphere}), 
\begin{eqnarray}
  \chi_0^i
  = \frac{\chi_0}{1 + 4\pi (L_i -1/3) \chi_0}
  .
  \label{eq:chi0_relation}
\end{eqnarray}
This equation shows that $\chi_0^i=\chi_0$ when $L_i=1/3$, as expected.

\section{General optical properties of ellipsoidal particle}
\label{sec:ellip}

The shape of an ellipsoidal particle is characterized by geometric factors \citep{Bohren+1983}:
\begin{eqnarray}
  L_i
  \equiv
  \frac{3 \mathcal{V}}{8 \pi} \int^\infty_0
  \frac{dq}{(q + a_i^2) \sqrt{(q + a_x^2) (q + a_y^2) (q + a_z^2)}} 
  ,
  \label{eq:Li}
\end{eqnarray}
where $i=x,\ y$ and $z$.
The volume of the ellipsoid is given by $\mathcal{V} = 4 \pi a_x a_y a_z / 3$.
The geometrical factors satisfy the following inequality:  $L_x \le L_y \le L_z$.
In addition, since these variables satisfy the identity  of $L_x + L_y + L_z = 1$, one of the three is not an independent variable.
We treat $L_x$ and $L_y$ as independent variables.
The continuous distributions of ellipsoids (CDE: \cite{Bohren+1983})  with a lower cut-off of $L_x$ at $L_\mathrm{min}$  is adopted as the shape parameter distribution.
This distribution is referred to the externally restricted CDE (ERCDE: \cite{Zubko+1996}).
A sphere is reproduced by setting $L_x = L_y = L_z = 1/3$.

The absorption cross sections of an ellipsoidal particle for radiation polarized along each axis are given by following equation:
\begin{eqnarray}
  C_{\nu,i}^\mathrm{abs}
  = \frac{8 \pi^2 \mathcal{V}}{\lambda} \mathrm{Im}
  \left( \chi_0^i \right)
  ,
  \label{eq:Cabs_i}
\end{eqnarray}
where $\chi_0^i$ is the complex susceptibility responding to an external electric field parallel to each axis (equation (\ref{eq:chi0_relation})).

The electric susceptibilities averaged over the shape distribution described by the ERCDE are deduced by \citet{Draine+2017} as,
\begin{eqnarray}
  \langle \chi_0^x \rangle
  &=&
  \frac{3}{2 \pi A^2 (\varepsilon - 1)}
  \left[ -A (\varepsilon - 1) + 3X \ln \left(\frac{X}{Y} \right) \right]
  ,
  \label{eq:chi0_x} \\
  \langle \chi_0^y \rangle
  &=&
  \frac{3}{\pi A^2 (\varepsilon - 1)}
  \left[ 3X \ln \left(\frac{Z}{X}\right)
  + Y \ln \left(\frac{Y}{Z}\right) \right]
  ,
  \label{eq:chi0_y} \\
  \langle \chi_0^z \rangle
  &=&
  \frac{3}{2 \pi A^2 (\varepsilon - 1)}
  \left[ 3X \ln \left(\frac{X}{Z}\right)
  + W \ln \left(\frac{W}{Z}\right) \right]
  ,
  \label{eq:chi0_z}
\end{eqnarray}
where
$A \equiv 1 - 3 L_\mathrm{min}$,
$X \equiv 1 + (\varepsilon - 1) / 3$,
$Y \equiv 1 + L_\mathrm{min} (\varepsilon - 1)$,
$Z \equiv 1 + (1 - L_\mathrm{min}) (\varepsilon - 1) / 2$, 
$W \equiv 1 + (1 - 2 L_\mathrm{min}) (\varepsilon - 1)$, and 
$\varepsilon$ is the electric permittivity of the amorphous dust.
The shape-averaged absorption cross sections for the electric field in the direction of each axis, $\langle C_{\nu,i}^\mathrm{abs} \rangle$, are derived by substituting equations (\ref{eq:chi0_x})--(\ref{eq:chi0_z}) for equation (\ref{eq:Cabs_i}). 

\bibliography{article}
\end{document}